\documentclass[a4paper,fleqn,usenatbib]{mnras}

\usepackage{newtxtext,newtxmath}
\usepackage[T1]{fontenc}
\usepackage{ae,aecompl}

\usepackage{graphicx}	% Including figure files
\usepackage{amsmath}	% Advanced maths commands
\usepackage{amssymb}	% Extra maths symbols
\usepackage{lscape}
\usepackage{rotating}
\usepackage{float}
\hypersetup{colorlinks = true, linkcolor = blue, urlcolor=blue, citecolor=blue} 
\usepackage[normalem]{ulem}
\usepackage{tabularx}
\usepackage{rotating, booktabs}
\usepackage{inputenc}
\usepackage{geometry}
\usepackage{changepage}
\usepackage{color}

\def\msun{\hbox{M$_\odot$}}
\def\vi{$m_{\rm F555W}-m_{\rm F814W} \,$}
\def\ub{$m_{\rm F336W}-m_{\rm F438W} \,$}
\def\b{$m_{\rm F438W} \, $}
\def\bi{$m_{\rm F438W}-m_{\rm F814W} \,$}
\def\i{$m_{\rm F814W} \, $}
\def\cubun{$C_{\rm F336W,F438W,F343N} \,$}
\def\v{$m_{\rm F555W} \, $}
\def\cunbi{$C_{\rm F343N,F438W,F814W} \,$}

\title[Cluster age and abundance spreads]{The search for multiple populations in Magellanic Clouds clusters - V. Correlation between cluster age and abundance spreads}

\author[Martocchia et al.]{S. Martocchia$^{1,2}$, 
E. Dalessandro$^{3}$, 
C. Lardo$^{4}$,
I. Cabrera-Ziri$^{5}$\thanks{Hubble Fellow.},
N. Bastian$^{2}$, 
\newauthor V. Kozhurina-Platais$^{6}$, 
M. Salaris$^{2}$,
W. Chantereau$^{2}$,
D. Geisler$^{7}$,
M. Hilker$^{1}$, 
\newauthor  N. Kacharov$^{8}$, 
S. Larsen$^{9}$,  
A. Mucciarelli$^{3, 10}$, 
F. Niederhofer$^{11}$,
I. Platais$^{12}$, 
C. Usher$^{2}$
\\
$^{1}$European Southern Observatory, Karl-Schwarzschild-Stra\ss e 2, D-85748 Garching bei M\"unchen, Germany\\
$^{2}$Astrophysics Research Institute, Liverpool John Moores University, 146 Brownlow Hill, Liverpool L3 5RF, UK\\
$^{3}$INAF-Osservatorio di Astrofisica \& Scienza dello Spazio, via Gobetti 93/3, I-40129, Bologna, Italy\\
$^{4}$Laboratoire d'astrophysique, \' Ecole Polytechnique F\' ed\' erale de Lausanne (EPFL), Observatoire, 1290, Versoix, Switzerland\\
$^{5}$Harvard-Smithsonian Center for Astrophysics, 60 Garden Street, Cambridge, MA 02138, USA\\
$^{6}$Space Telescope Science Institute, 3700 San Martin Drive, Baltimore, MD 21218, USA\\
$^{7}$Departamento de Astronomia, Universidad de Concepcion, Casilla 160-C, Chile\\
$^{8}$Max-Planck-Institut f\"ur Astronomie, K\"onigstuhl 17, D-69117 Heidelberg, Germany\\
$^{9}$Department of Astrophysics/IMAPP, Radboud University, P.O. Box 9010, 6500 GL Nijmegen, The Netherlands\\
$^{10}$Dipartimento di Fisica \& Astronomia, Universit\` a degli Studi di Bologna, via Gobetti 93/2, I-40129, Bologna, Italy\\
$^{11}$Leibniz-Institut f\"ur Astrophysik Potsdam, An der Sternwarte 16, Potsdam 14482, Germany\\
$^{12}$Department of Physics and Astronomy, Johns Hopkins University, 3400 North Charles Street, Baltimore, MD 21218, USA\\}

\date{Accepted XXX. Received YYY; in original form ZZZ}

\pubyear{2019}

\begin{document}
\label{firstpage}
\pagerange{\pageref{firstpage}--\pageref{lastpage}}
\maketitle

% Abstract of the paper
\begin{abstract}
In our HST photometric survey, we have been searching for multiple stellar populations (MPs) in Magellanic Clouds (MCs) massive star clusters which span a significant range of ages ($\sim 1.5-11$ Gyr). In the previous papers of the series, we have shown that the age of the cluster represents one of the key factors in shaping the origin of the chemical anomalies. 
Here we present the analysis of four additional clusters in the MCs, namely Lindsay 38, Lindsay 113, NGC 2121 and NGC 2155, for which we recently obtained new UV HST observations. These clusters are more massive than $\sim 10^4$ \msun \, and have ages between $\sim 2.5-6$ Gyr, i.e. located in a previously unexplored region of the cluster age/mass diagram. We found chemical anomalies, in the form of N spreads, in three out of four clusters in the sample, namely in NGC 2121, NGC 2155 and Lindsay 113. 
By combining data from our survey and HST photometry for 3 additional clusters in the Milky Way (namely 47 Tuc, M15 and NGC 2419), we show that the extent of the MPs in the form of N spread is a strong function of age, with older clusters having larger N spreads with respect to the younger ones.
Hence, we confirm that cluster age 
plays a significant role in the onset of MPs.  
\end{abstract}

% Select between one and six entries from the list of approved keywords.
% Don't make up new ones.

%%%%%%%%%%%%%%%%%%%%%%%%%%%%%%%%%%%%%%%%%%%%%%%%%%

%%%%%%%%%%%%%%%%% BODY OF PAPER %%%%%%%%%%%%%%%%%%

\begin{table*}
\caption{Log of the HST observations used in this paper.}
\label{tab:log}
\centering          
    \begin{tabular}{c c c c c c}    
        \toprule\toprule
        Cluster Name & GO & Camera & Filter & N $\times$ exp. time & P.I. \\
        \midrule
        NGC 2121 & 15062 & WFC3/UVIS & F336W & 2$\times$715 s, 270 s & N. Bastian \\
                        & 15062 & WFC3/UVIS & F343N & 2$\times$1060 s, 540 s & N. Bastian\\
                        & 15062 & WFC3/UVIS & F438W & 2$\times$550 s, 120 s & N. Bastian\\
                        &  8141  & WFPC2       & F555W & 4$\times$400 s & R. Rich \\
                        &  8141  & WFPC2       & F814W & 4$\times$400 s & R. Rich \\
        NGC 2155 & 15062 & WFC3/UVIS & F336W & 2$\times$705 s, 250 s & N. Bastian \\
                        & 15062 & WFC3/UVIS & F343N & 2$\times$1060 s, 530 s & N. Bastian\\
                        & 15062 & WFC3/UVIS & F438W & 2$\times$545 s, 120 s & N. Bastian\\
                        & 5475  & WFPC2         & F450W & 230 s & M. Shara \\
                        & 5475  & WFPC2         & F555W & 120 s & M. Shara \\
        Lindsay 38 & 15062 & WFC3/UVIS & F336W & 2$\times$710 s, 268 s & N. Bastian \\
                        & 15062 & WFC3/UVIS & F343N & 2$\times$1057 s, 515 s & N. Bastian\\
                        & 15062 & WFC3/UVIS & F438W & 2$\times$538 s, 123 s & N. Bastian\\
                        & 10396 & ACS/WFC    & F555W & 4$\times$485 s, 2$\times$20 s & J. Gallagher \\
                        & 10396 & ACS/WFC    & F814W & 4$\times$463 s, 2$\times$10 s & J. Gallagher \\
        Lindsay 113 & 15062 & WFC3/UVIS & F336W & 2$\times$720 s, 274 s & N. Bastian\\
                        & 15062 & WFC3/UVIS & F343N & 2$\times$1065 s, 530 s & N. Bastian\\
                        & 15062 & WFC3/UVIS & F438W & 2$\times$545 s, 128 s & N. Bastian\\
                        & 9891  & ACS/WFC    & F555W & 480 s & G. Gilmore \\
                        & 9891  & ACS/WFC    & F814W & 290 s &  G. Gilmore \\
       \bottomrule\bottomrule
        \end{tabular}
\end{table*}

\section{Introduction}
\label{sec:intro}

It is now well established that globular clusters (GCs) host star-to-star light element abundance variations, which are typically referred to as multiple populations (MPs).  
Several scenarios have been proposed over the years to explain the formation and observed properties of MPs, however their origin is still unclear and strongly debated in the literature (e.g., \citealt{renzini15}, \citealt{aarev18}). 

Until a few years ago, these chemical variations had only been found 
in massive clusters older than $\sim$10 Gyr. 
Regardless of environment, MPs have been discovered in almost all ancient clusters surveyed 
in the Milky Way (MW, \citealt{gratton12}), Magellanic Clouds (MCs, \citealt{mucciarelli09, dalessandro16, niederhofer17a,gilligan19}), Fornax Dwarf Galaxy \citep{larsen14} and the Sagittarius dwarf galaxy (e.g. M54, \citealt{carretta10}).
The absence of light-element variations was suggested in a handful of massive and old Galactic GCs (e.g., \citealt{walker11,villanova13}). However, recent detailed studies have demonstrated that MPs are indeed  present also in these systems (see for example the cases of IC 4499 and Rup 106; \citealt{dalessandro18,dotter18}).

We are conducting a joint Hubble Space Telescope (HST) and Very Large Telescope (VLT) survey with the goal of pinpointing the main physical mechanisms at the basis of MP formation. We targeted star clusters which are as massive as old GCs ($>$ a few times $10^4$ \msun), but span a wide range of ages (from $\sim$ 1.5 up to 11 Gyr). 
In total we targeted nine star clusters in the MCs in our HST photometric survey while four clusters were targeted in our VLT spectroscopic survey, with two targets in common between the surveys. 

We find that, together with mass \citep{carretta10,bragaglia12,schiavon13,milone17}, cluster age plays a key role in defining the onset and properties of chemical anomalies. In fact, we detect (within our photometric errors) MPs only in clusters older than $\sim2$ Gyr, with NGC 1978 (\citealt{martocchia18a}) and Hodge 6 \citep{hollyhead19} being the youngest systems where chemical variations have been detected to date.  

We stress we refer here to clusters with light-element star-to-star variations. While the colour-magnitude diagrams (CMDs) of young clusters ($<2$ Gyr) show multiple and extended main sequences (extended main sequence turnoffs, eMSTOs, e.g. \citealt{mackey08,milone09,bastian16}) which are often referred to as ``multiple populations'', it appears that the two phenomena are not directly related
%, as these young clusters do not show light element variations in their RGB 
(e.g., \citealt{mucciarelli14,martocchia17,martocchia18a}).  Instead, the observed complexities in the younger clusters are most likely caused by changes in the stellar structure of stars, caused by, for example, stellar rotation (e.g., \citealt{bastiandemink09,dantona15,milone18,kamann18,bastian18}). 
%Also, \cite{martocchia18b} report that NGC 1978 do not show an eMSTO feature, as it is expected from predictions from the stellar rotation scenario. Interestingly, the eMSTO phenomenon seems to disappear at the age where MPs seem to originate ($\sim 2$ Gyr).

So far, our initial sample had a gap between 2 and 6 Gyr and it also focussed on clusters with mass $\gtrsim 10^5$\msun.
In the current paper, we present a photometric study and search for MPs in four additional clusters in the MCs, namely NGC 2121, NGC 2155, Lindsay 38 and Lindsay 113, for which we recently obtained new HST UV observations. These clusters were chosen to sample the parameter space missed in our previous observations: they have ages between 
$\sim$2.5 and $\sim$6 Gyr and masses $M\lesssim10^5$ \msun.
We constrained the presence and amplitude of N abundance variations by analysing their RGB widths, consistently with what was done in \cite{martocchia18a}. 
Results are compared with what was obtained for the other clusters of the survey (\citealt{niederhofer17a,niederhofer17b,martocchia17,martocchia18a}, hereafter Papers I, II, III, IV) and for Galactic GCs (namely 47 Tuc, M15 and NGC 2419).

%We analysed the width of their red giant branches (RGBs) in the UV CMDs to look for a significant broadening with respect to the photometric errors, and establish whether MPs in the form of N spread are present in such clusters. We also compared the widths of the RGBs with the other clusters in our sample and with three additional GCs in the MW (namely ), for which we used archival HST photometry. 

This paper is organised as follows: in \S \ref{sec:obs} we describe the photometric reduction procedures, while we report on
the analysis used to quantify the detection of MPs in \S \ref{sec:analysis}. In \S \ref{sec:results} we present the main results of the paper and we compare all the clusters in our HST survey.
We finally discuss and conclude in \S \ref{sec:disc}.

\section{Observations and data reduction}
\label{sec:obs}

The observations of the four clusters analysed in this paper are from the \textit{Hubble Space Telescope} (HST) and they consist in both proprietary and archival data. All clusters were observed through proposal GO-15062 (PI: N. Bastian) with the WFC3/UVIS camera in the F336W, F343N and F438W filters.
These observations were then complemented with different archival
data. In particular, for NGC 2121, we used archival WFPC2 observations from the program GO-8141 (P.I. R. Rich), taken with the F555W and F814W filters. 
For NGC 2155, we also used archival WFPC2 observations in the F450W and F555W filters, program ID GO-5475 (P.I. M. Shara). We complemented the observations for Lindsay 38 with archival ACS data in F555W and F814W filters (program ID GO-10396, P.I. J. Gallagher) and for Lindsay 113 we also used F555W and F814W filters of the ACS instrument obtained in the program GO-9891, P.I. G. Gilmore). Table \ref{tab:log} provides information about the HST observations used in this paper. 

The images have been processed, flat-field corrected, and bias-subtracted by using standard HST pipelines ($flc$ images for WFC3/ACS and $c0f$ images for WFPC2). Pixel-area effects have been corrected by applying the Pixel Area Maps images to each WFC3/ACS image. We also corrected all images for cosmic rays contamination by using the L.A. Cosmic algorithm \citep{vandokkum01}.

The photometric analysis has been performed following the same strategy as in \cite{dalessandro14, dalessandro18}. Briefly, we used DAOPHOTIV \citep{stetson87} independently on each camera and each chip. We selected several hundreds of bright and isolated stars in order to model the point-spread function (PSF). All available analytic functions were considered for the PSF fitting (Gauss, Moffat, Lorentz and Penny functions), leaving the PSF free to spatially vary to the first-order. In each image, we then fit all the star-like sources detected at $3\sigma$ from the local background with the best-fit PSF model by using ALLSTAR. We then created a master catalogue composed of stars detected in (n/2 +1) images for each cluster\footnote{Where the number of exposures in the same filter is equal to three, we used stars detected in 2 images to create the catalogues.}. At the corresponding positions of stars in this final master-list, a fit was forced with DAOPHOT/ALLFRAME \citep{stetson94} in each frame. For each star thus recovered, multiple magnitude estimates obtained in each chip were homogenised by using DAOMATCH and DAOMASTER, and their weighted mean and standard deviation were finally adopted as star magnitude and photometric error. The final result consists in a catalogue for each camera\footnote{As an additional check we repeated the photometric analysis by using a third-order spatial variation for the PSF. However, we decided to perform the analysis on the catalogue where the PSF was left free to spatially vary to the first-order. No significant changes were detected between the two catalogues.}.

Instrumental magnitudes have been converted to the VEGAMAG photometric system by using the prescriptions and zero-points reported on the dedicated HST web-pages\footnote{see \url{http://www.stsci.edu/hst/wfc3/phot_zp_lbn} and \url{http://www.stsci.edu/hst/acs/analysis/zeropoints}}. Instrumental coordinates were reported on the absolute image World Coordinate System by using CataXcorr\footnote{Part of a package of astronomical softwares (CataPack) developed by P. Montegriffo at INAF-OABo.}. The WFC3 catalogue was combined with the ACS (or WFPC2) by using the same CataXcorr and CataComb.

Regarding the galactic GCs, the catalogue for NGC 2419 is from \cite{larsen19} which comprises WFC3 observations for F438W, F555W and F814W from GO-11903 (P.I. J. Kalirai) and UV data in the F336W and F343N bands from GO-15078 (P.I. S. Larsen). Data for M15 in the F343N and F438W bands are from GO-13295 (P.I. S. Larsen), which were cross-matched with the catalogue from the HUGS (HST UV Globular clusters Survey, \citealt{piotto15,nardiello18}) comprising observations in the F336W and F814W bands. HST F343N and F438W/WFC3 observations for 47 Tuc are from GO-14069 (P.I. N. Bastian), while F336W data are from GO-11729 (P.I. J. Holtzman). The optical filters F555W and F814W are from ACS GO-9443 (P.I. I. King).
These data, along with their analysis and comparison with the current dataset as well as with the HST UV Legacy Survey, will be discussed in detail in Cabrera-Ziri et al. (in preparation).

\begin{figure}
\centering
\includegraphics[scale=0.38]{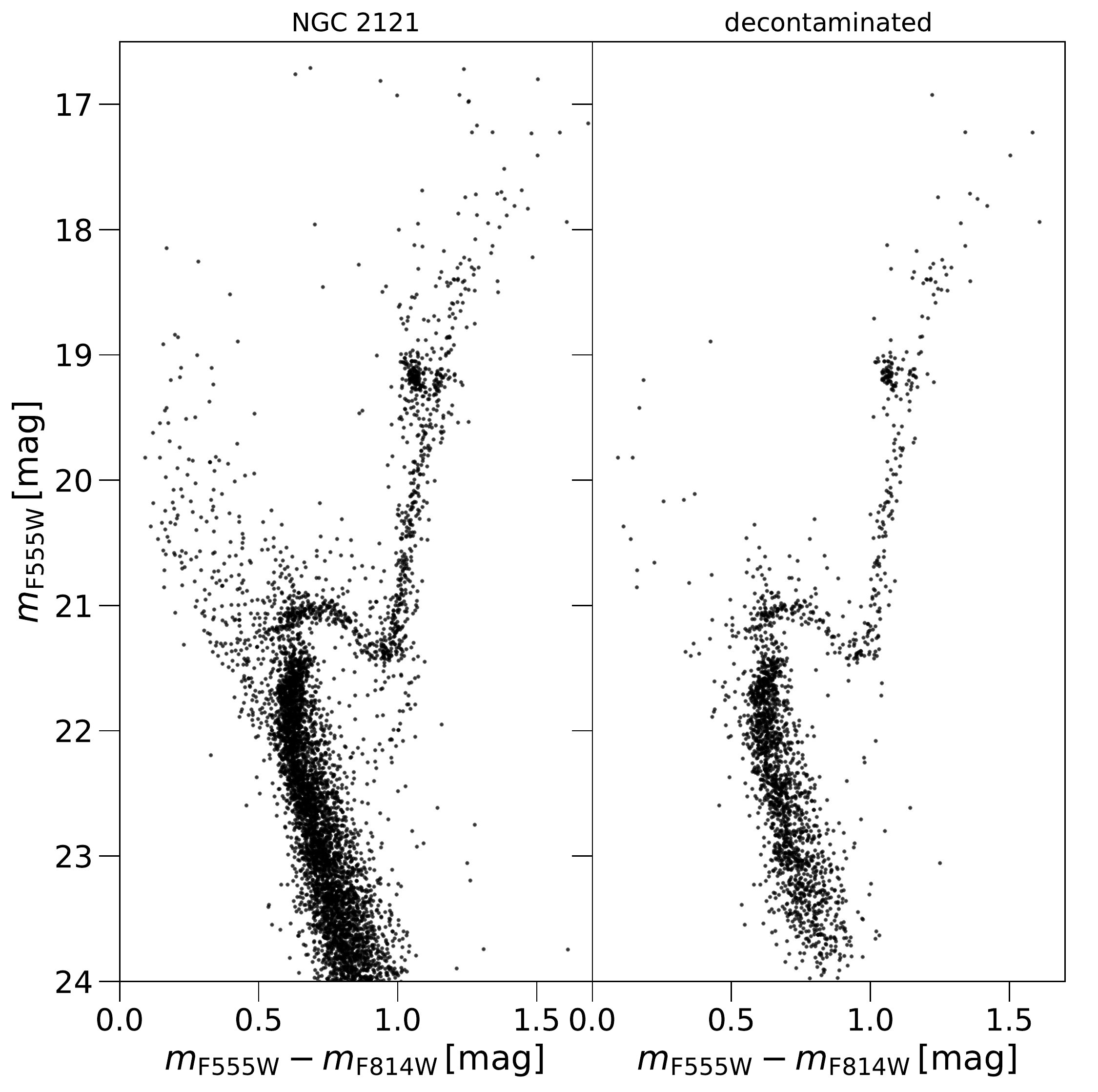}
\caption{\vi vs. \v CMD of NGC 2121 before (left panel) and after (right panel) the field star subtraction.} 
\label{fig:2121fssVI}
\end{figure}

\subsection{Artificial Star Tests}
\label{subsec:AS}

We performed artificial star (AS) experiments for each cluster following the method described in \cite{dalessandro15} (see also \citealt{bellazzini02,dalessandro16}) to derive a reliable estimate of the photometric errors. 

We generated a catalog of simulated stars with an input magnitude extracted from a luminosity function (LF) modeled to reproduce the observed LF in that band and extrapolated beyond the observed limiting magnitude. We then assigned an input magnitude for each filter involved 
to each star, extracting it from the luminosity function, by means of an interpolation along the ridge mean lines that were obtained in different CMDs by averaging over 0.4 mag bins and applying a $2\sigma$ clipping algorithm.

\begin{figure*}
\centering
\includegraphics[scale=0.38]{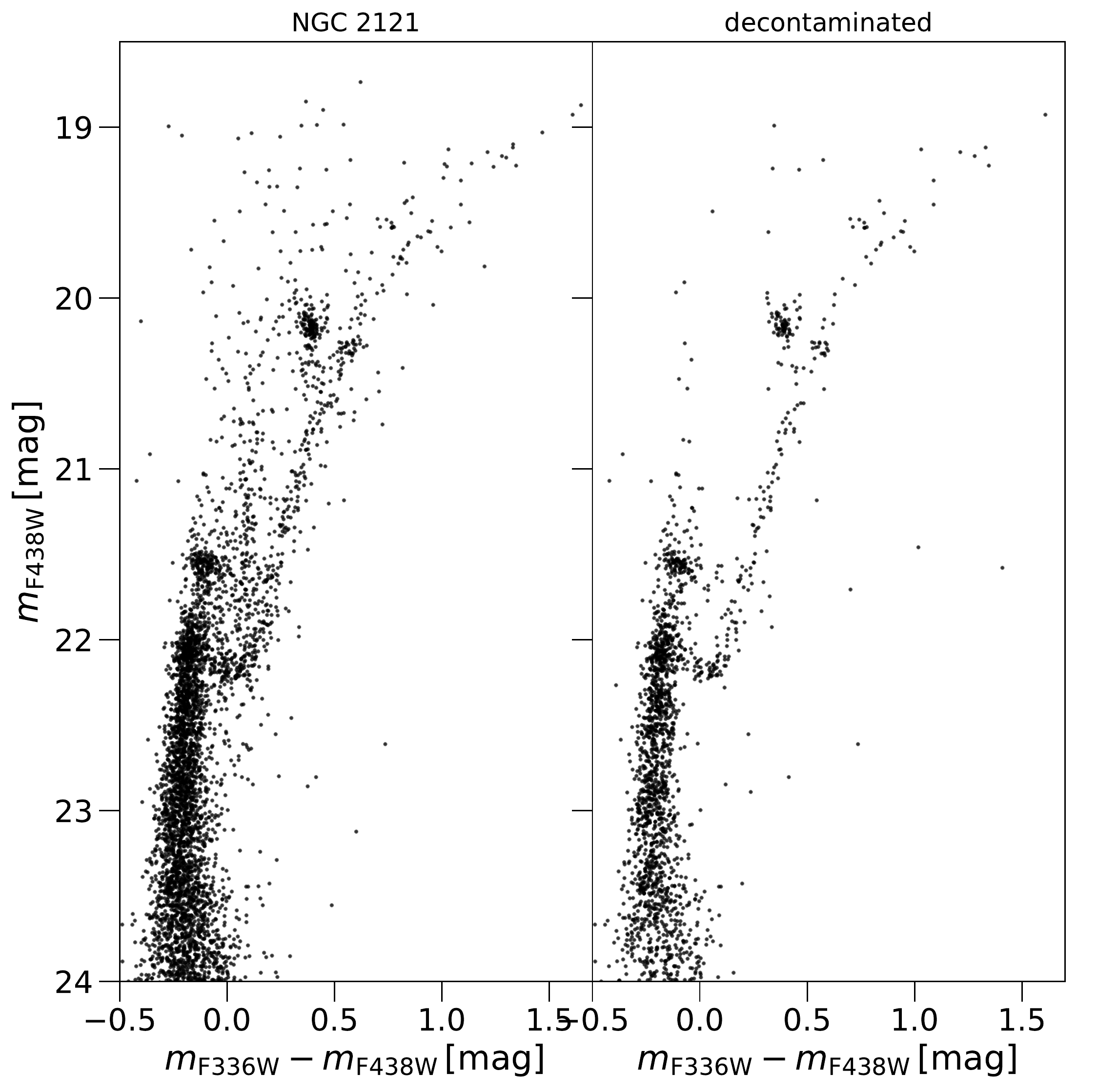}
\includegraphics[scale=0.38]{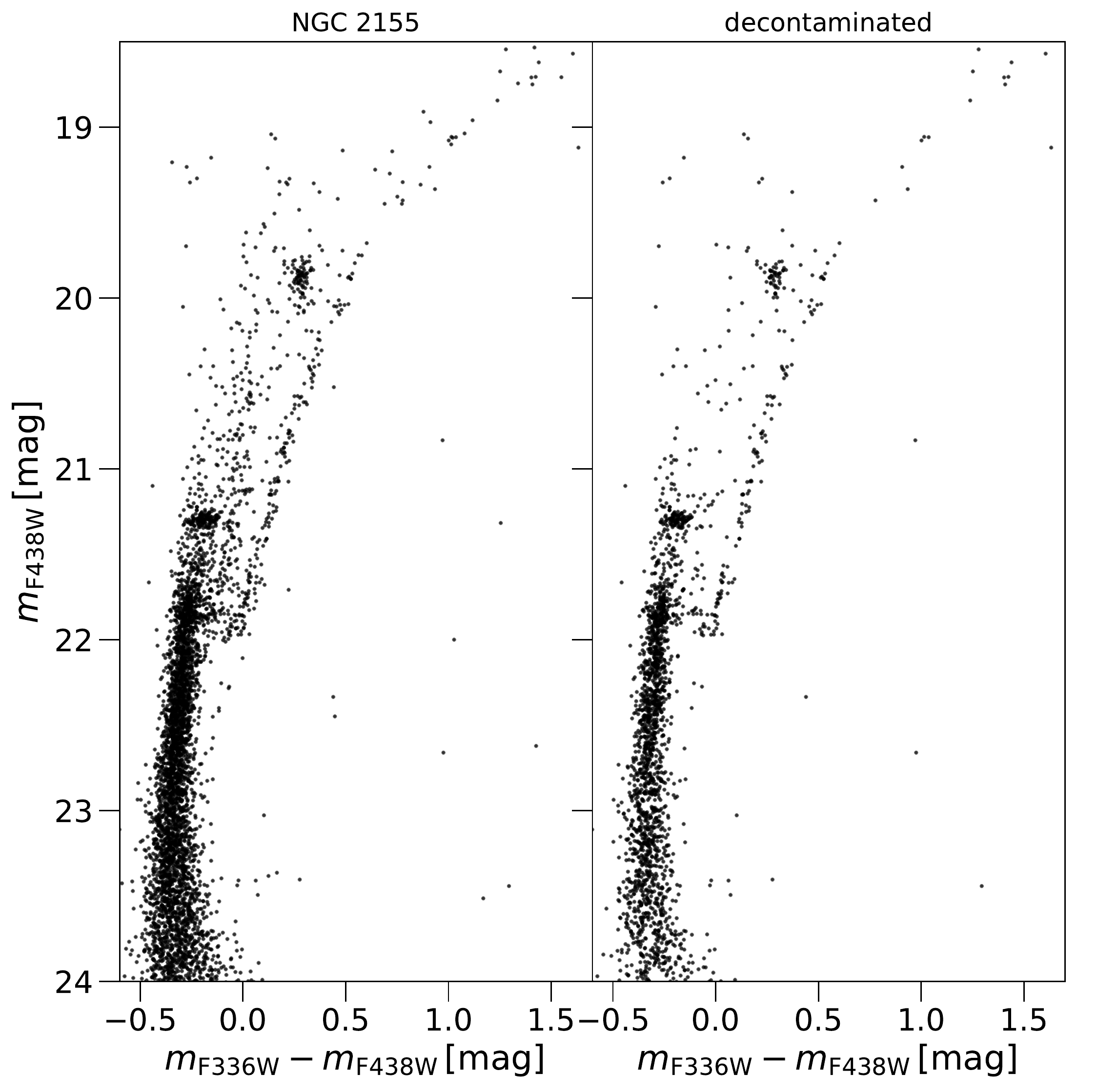}
\caption{\ub vs. \b CMDs of NGC 2121 (left panels) and NGC 2155 (right panels) before and after the field star subtraction.} 
\label{fig:fssUB}
\end{figure*}

Artificial stars were added to real images (which include also real stars) by using the software {\tt DAOPHOTII/ADDSTAR} \citep{stetson87}.
Then, the photometric analysis was performed using the same reduction strategy and PSF models used for real images (see above for details) on both real and simulated stars. In this way, the effect of radial variation of crowding on both completeness and photometric errors is accounted for. Artificial crowding was minimized by placing stars into the images following a regular grid composed by $15\times 15$ pixel cells in which only one artificial star for each run was allowed to lie at a random position within the cell. 
For each run, we simulated in this way $\sim 14,000$ stars. After a large number of experiments, stars are uniformly distributed in coordinates. The procedure was repeated until a minimum number of 50,000 artificial stars was added to each ACS/WFC3/WFPC2 chip.

\begin{figure}
\centering
\includegraphics[scale=0.4]{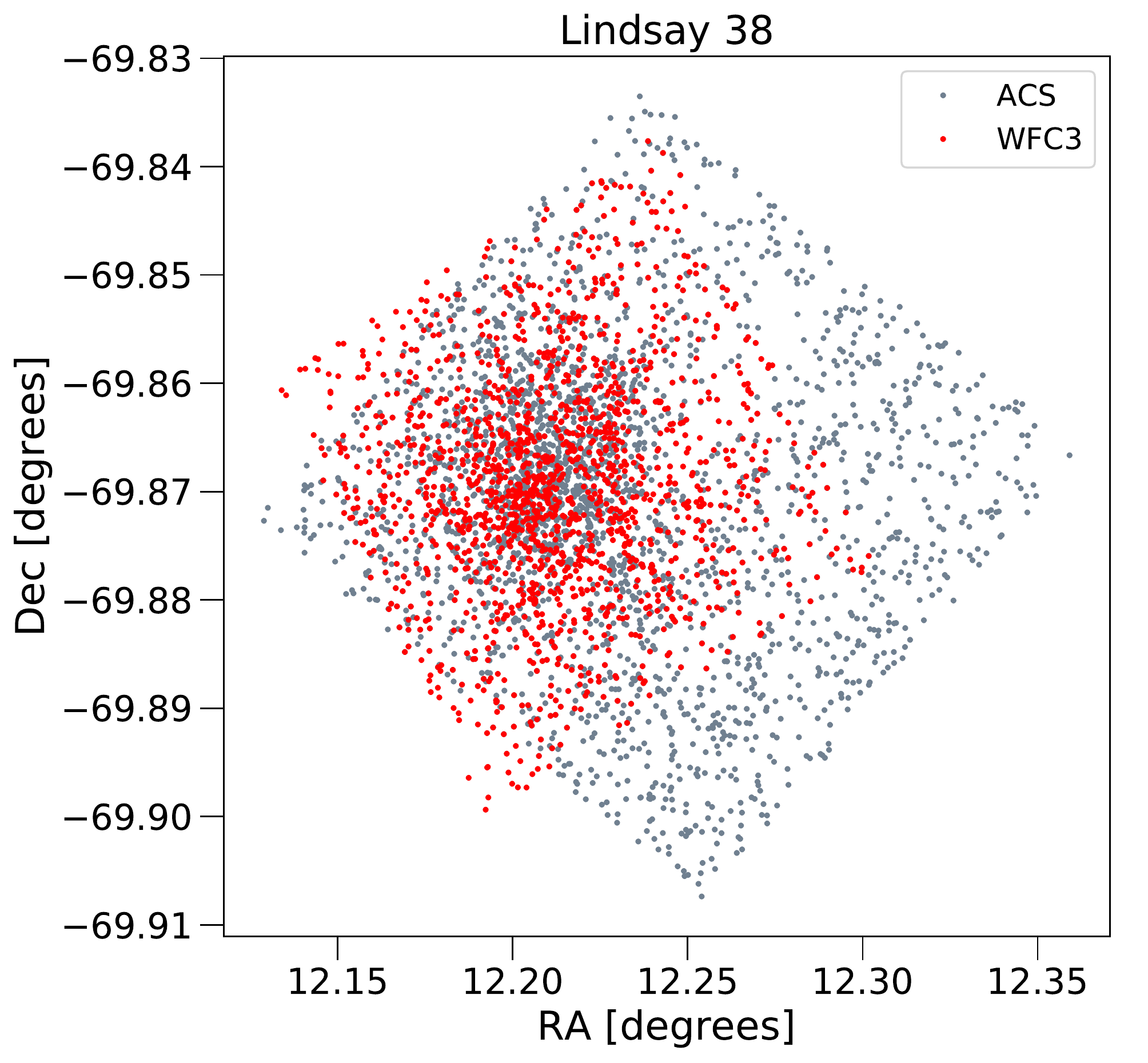}
\includegraphics[scale=0.4]{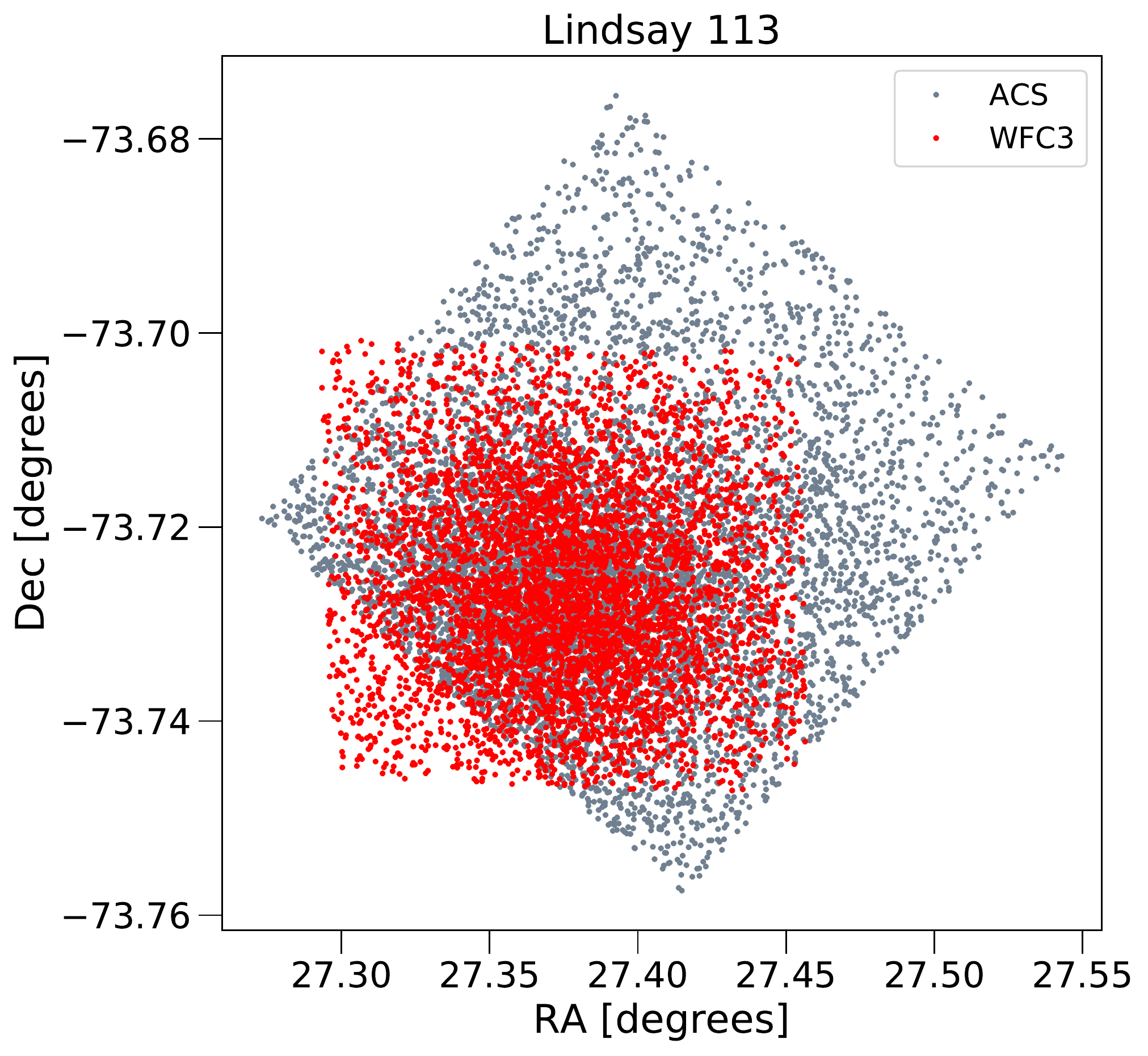}
\caption{ACS(grey) and WFC3(red) map for Lindsay 38 (top) and Lindsay 113 (bottom) FOV.} 
\label{fig:l38_fov}
\end{figure}

\section{Analysis}
\label{sec:analysis}

%We estimated the centre of each cluster by 
%Then, we fit a King profile \citep{king62} to the stellar density to estimate the radial profile of each cluster. The stellar density was calculated in concentric annuli until a maximum radius which is different from cluster to cluster depending on the field of view(FOV) available. 
For NGC 2121, NGC 2155 and Lindsay 113, the analysis presented in this paper was performed in a circular region around the cluster centre (the ``cluster region''). Stars were selected within a radius of 40 arcsec from the centre of NGC 2121 and NGC 2155 and within a radius of 45 arcsec from the centre of Lindsay 113\footnote{The radii of the clusters were selected to maximise the areas of the background and cluster regions at the same time, avoiding overlap between the two.}. The centre of each cluster was estimated by fitting a two-dimensional Gaussian to the distribution of the stellar density. For NGC 2121 and NGC 2155, we performed a statistical decontamination analysis to use likely cluster members. The background subtraction technique was extensively discussed in our previous papers, thus we refer the interested reader to Paper I to IV for more details. Figure \ref{fig:2121fssVI} shows the \vi vs. \v CMD of NGC 2121 before (left panel) and after (right panel) the field star subtraction, while Fig. \ref{fig:fssUB} reports the \ub vs. \b CMDs of NGC 2121 and NGC 2155 before and after the decontamination. 

\begin{figure*}
\centering
\includegraphics[scale=0.33]{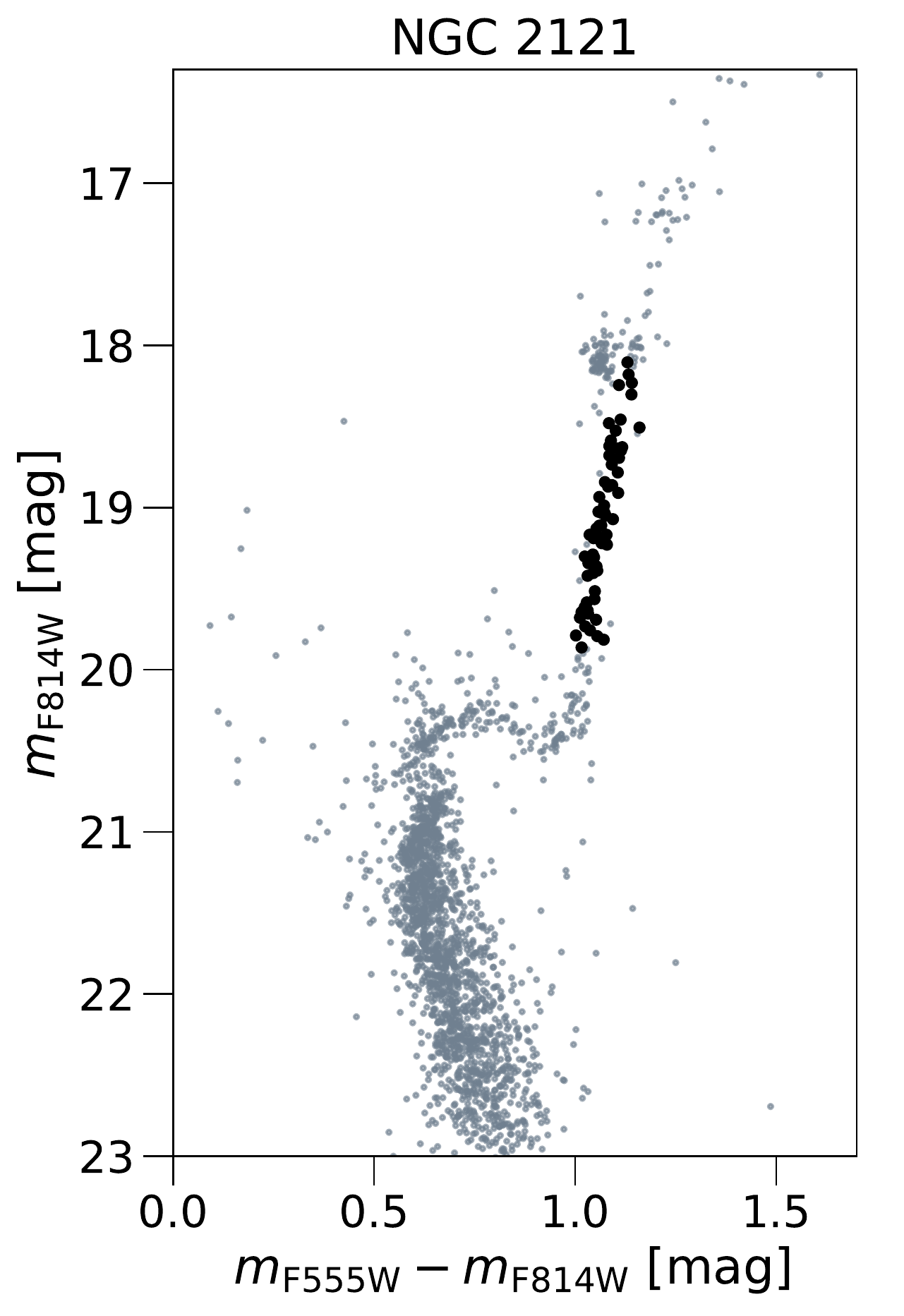}
\includegraphics[scale=0.33]{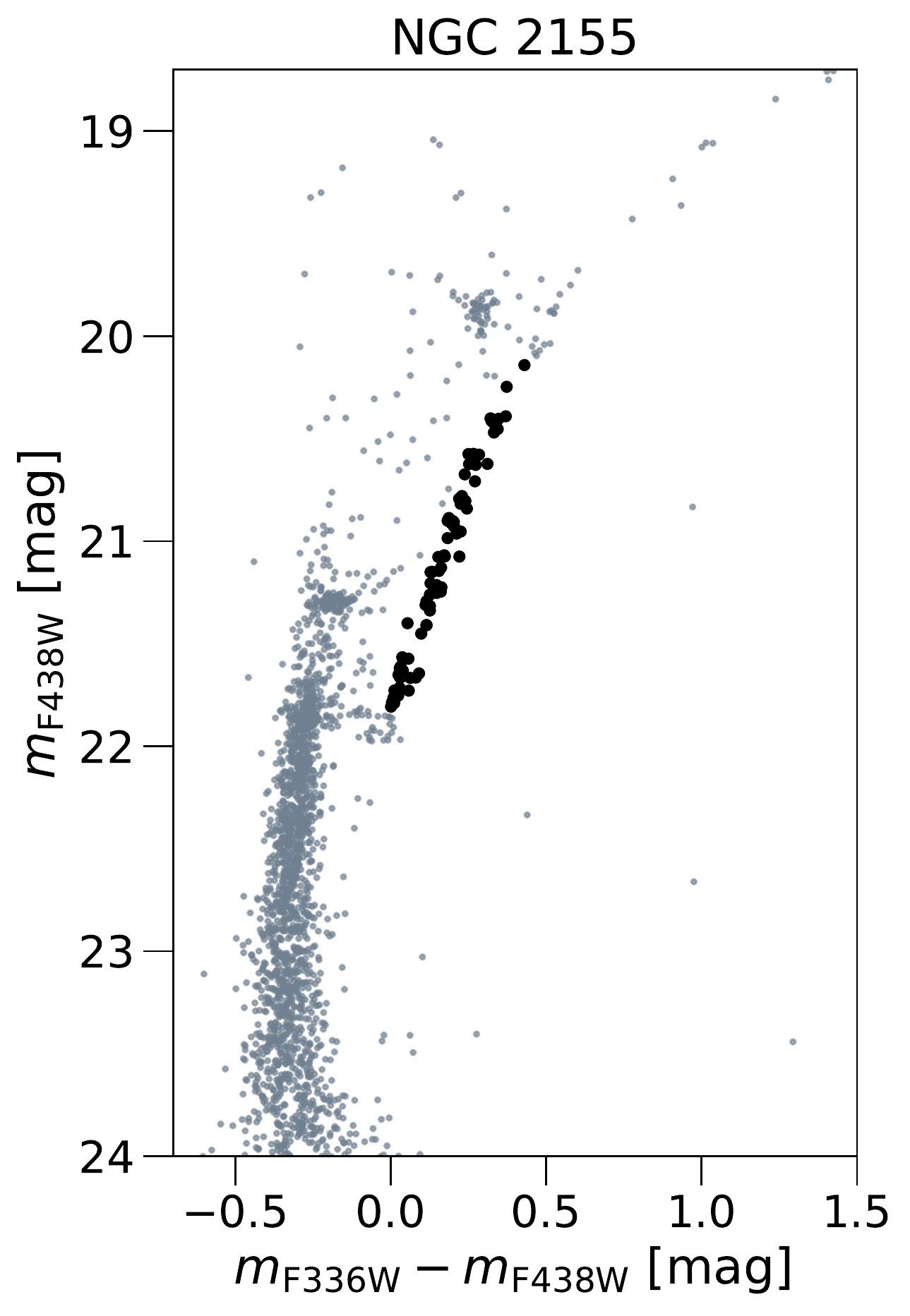}
\includegraphics[scale=0.33]{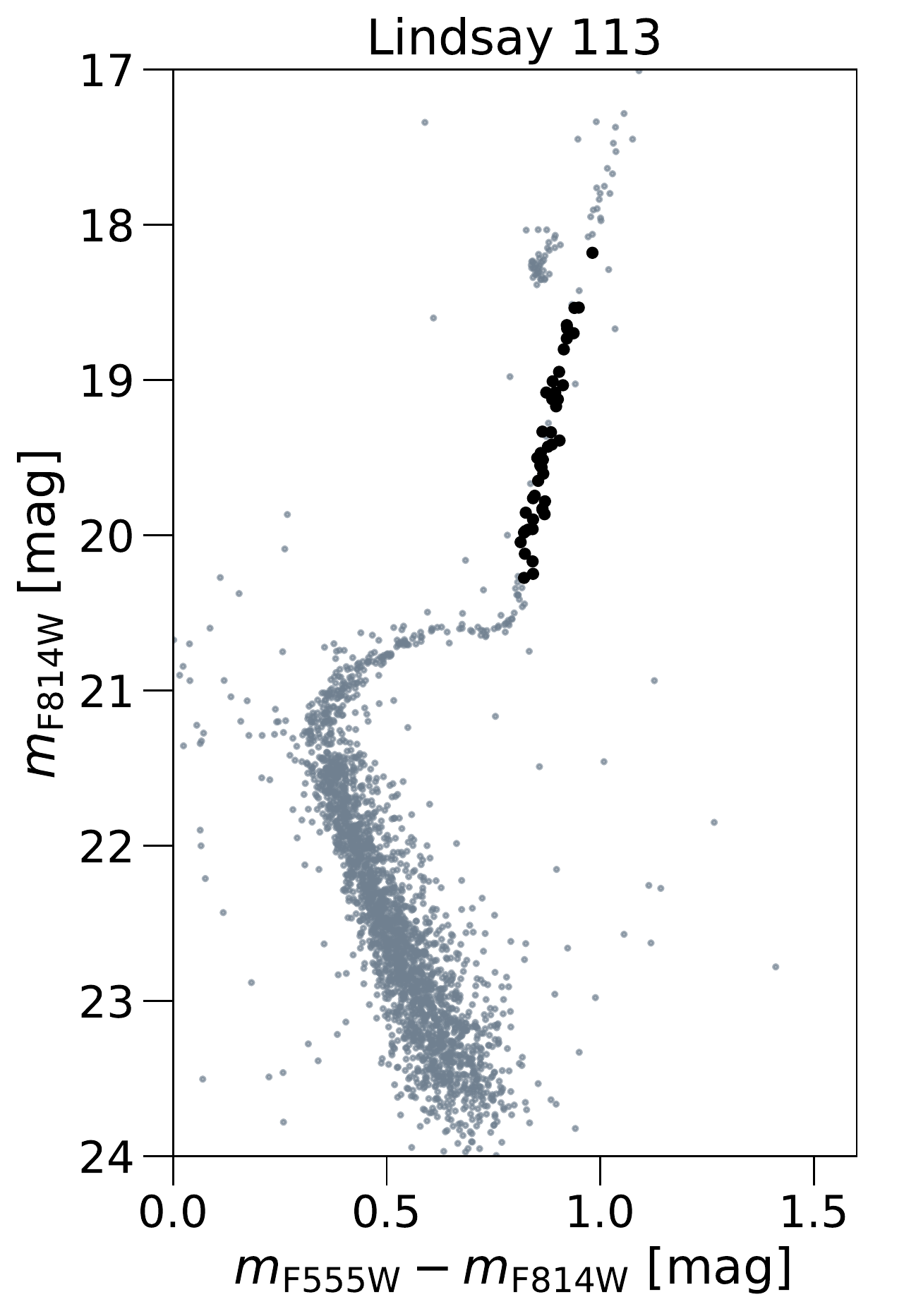}
\includegraphics[scale=0.33]{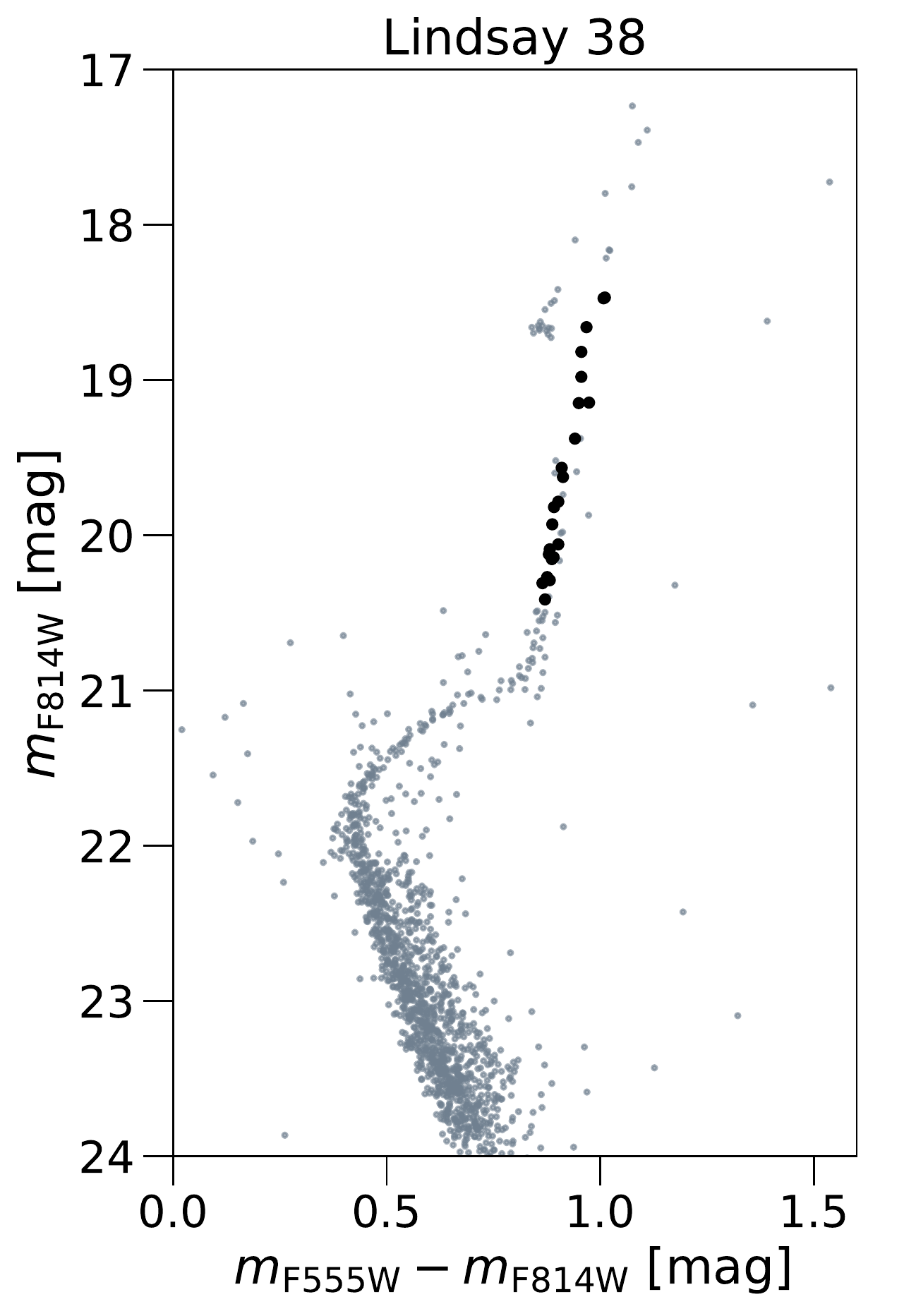}
\caption{\vi vs. \i CMDs for NGC 2121, Lindsay 113, Lindsay 38 and \ub vs. \b CMD for NGC 2155. Black circles indicate the final selected RGB stars.} 
\label{fig:viRGB}
\end{figure*}

Field stars were not subtracted in Lindsay 38 and Lindsay 113 since 
it was not possible to define a background region that is extended enough. Figure \ref{fig:l38_fov} shows the RA vs. Dec map for Lindsay 38 (top) and Lindsay 113 (bottom), where the FOV covered by the ACS(WFC3) camera is indicated in grey(red). A first look reveals that the \vi vs. \v CMDs of Lindsay 38 and Lindsay 113 are rather clean from field stars contamination (see Fig. \ref{fig:viRGB}). In the current analysis, we considered all the stars in common between the ACS and WFC3 catalogue for Lindsay 38.

We note that, for NGC 2155, the WFPC2 catalogue was only used to give an estimate of the age of the cluster (see \S \ref{subsec:age}). The optical images (F450W and F555W/WFPC2) only have one exposure per filter (Table \ref{tab:log}) and we found that the addition of these filters to the catalogue was not useful, instead it only added noise.

We also corrected our photometric catalogues for differential reddening by using the same method reported in \cite{milone12} and \cite{dalessandro18}. We found that our clusters are not significantly affected by differential reddening, with a maximum $\delta E(B-V)$ of $\sim$0.003 mag for Lindsay 38 and $\sim$0.005 mag for Lindsay 113 and NGC 2121. 

As in our previous analysis (Papers I to IV), we first selected bona-fide RGB stars in the \vi vs \i CMD and then in the \bi vs \i CMD, except for NGC 2155 where RGB stars were selected in the \ub vs \b CMD. 
Stars were selected between the base of the lower RGB ($\sim 0.5$ magnitude above the main sequence turnoff) and the RGB bump, to avoid contamination by SGB or AGB stars.
Figure \ref{fig:viRGB} shows the \vi vs \i CMDs of NGC 2121, Lindsay 113 and Lindsay 38, and the \ub vs. \b CMD for NGC 2155. Black filled circles indicate the final  selected RGB stars.

\begin{figure*}
\centering
\includegraphics[scale=0.38]{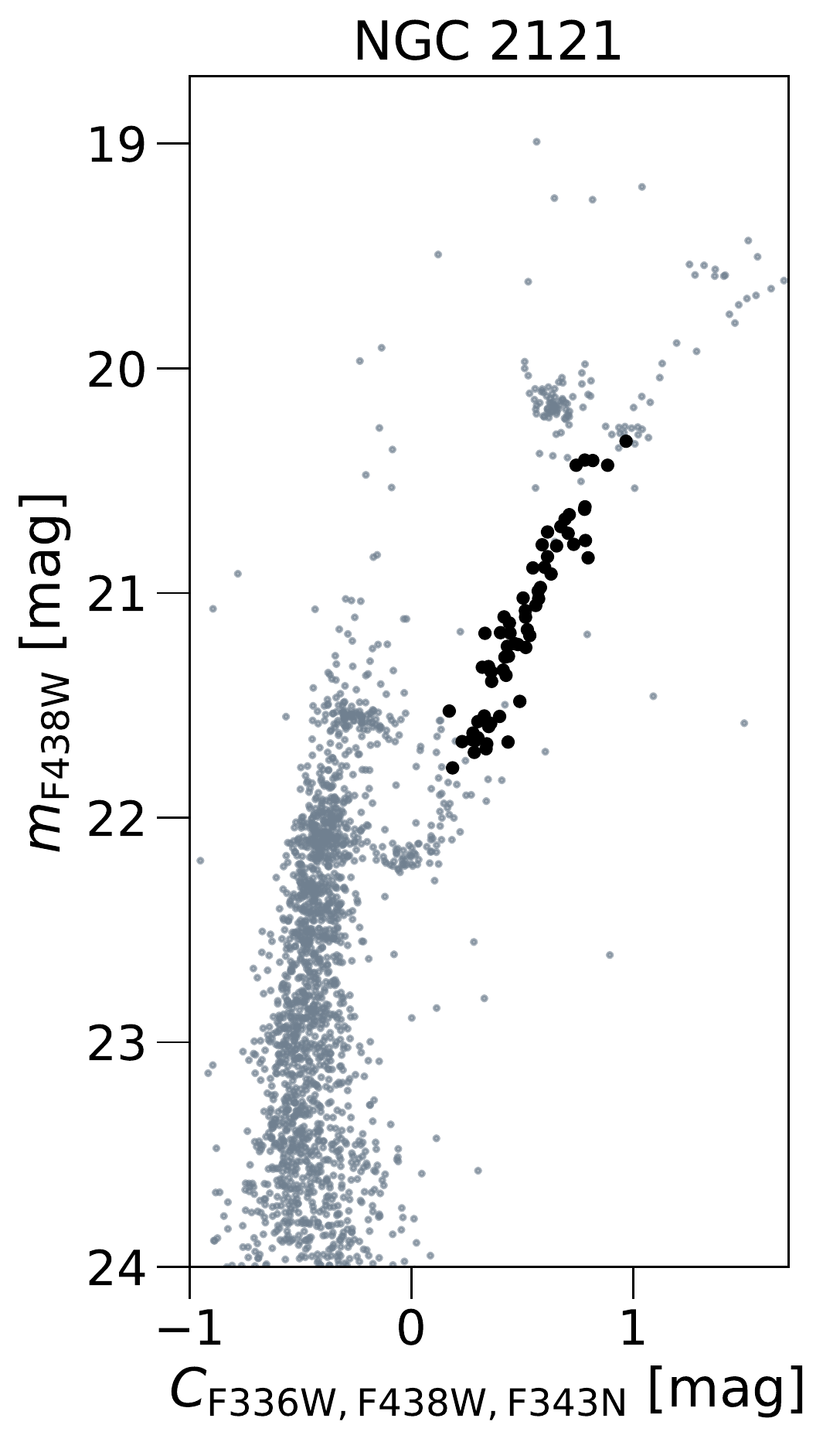}
\includegraphics[scale=0.38]{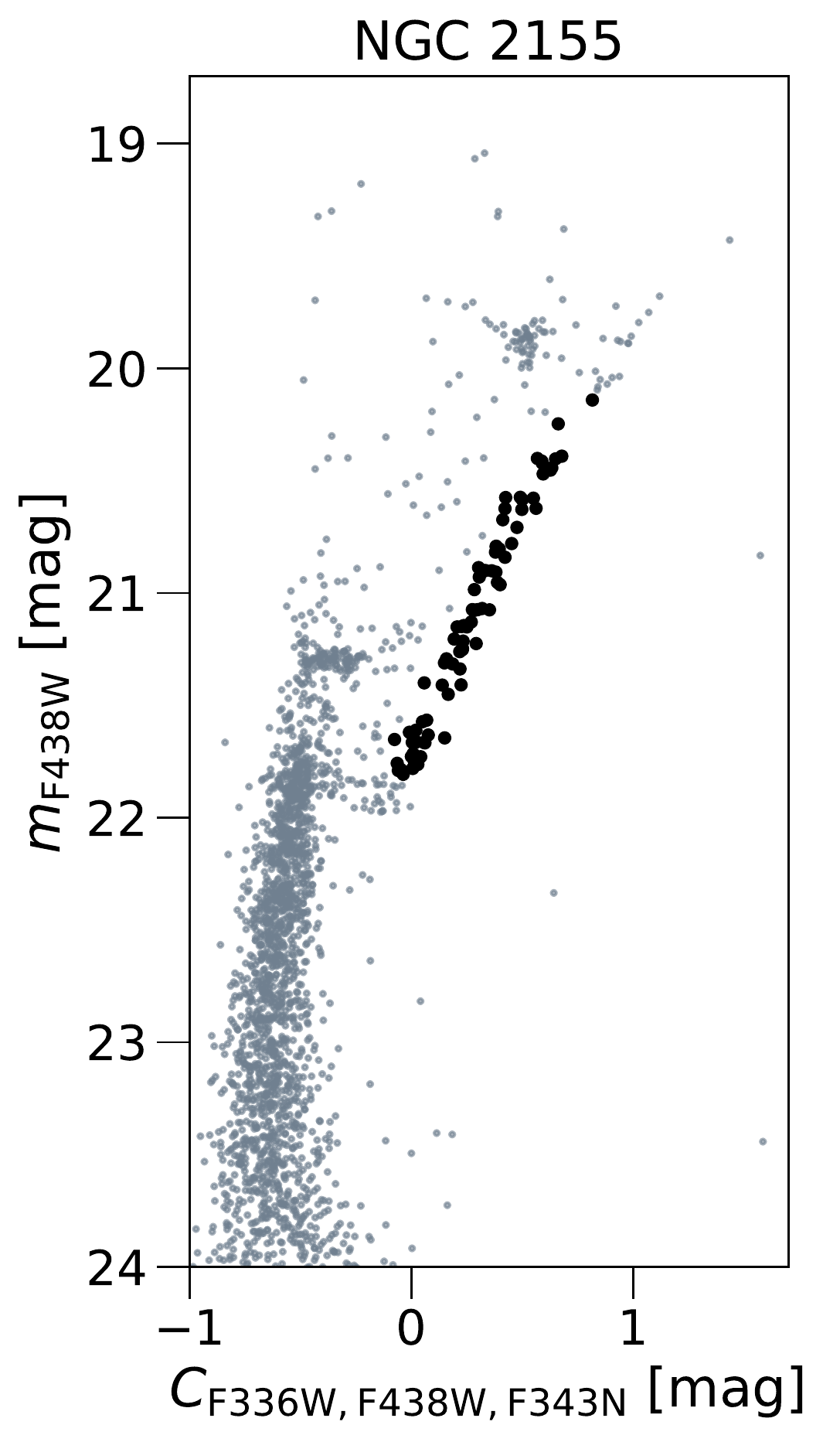}
\includegraphics[scale=0.38]{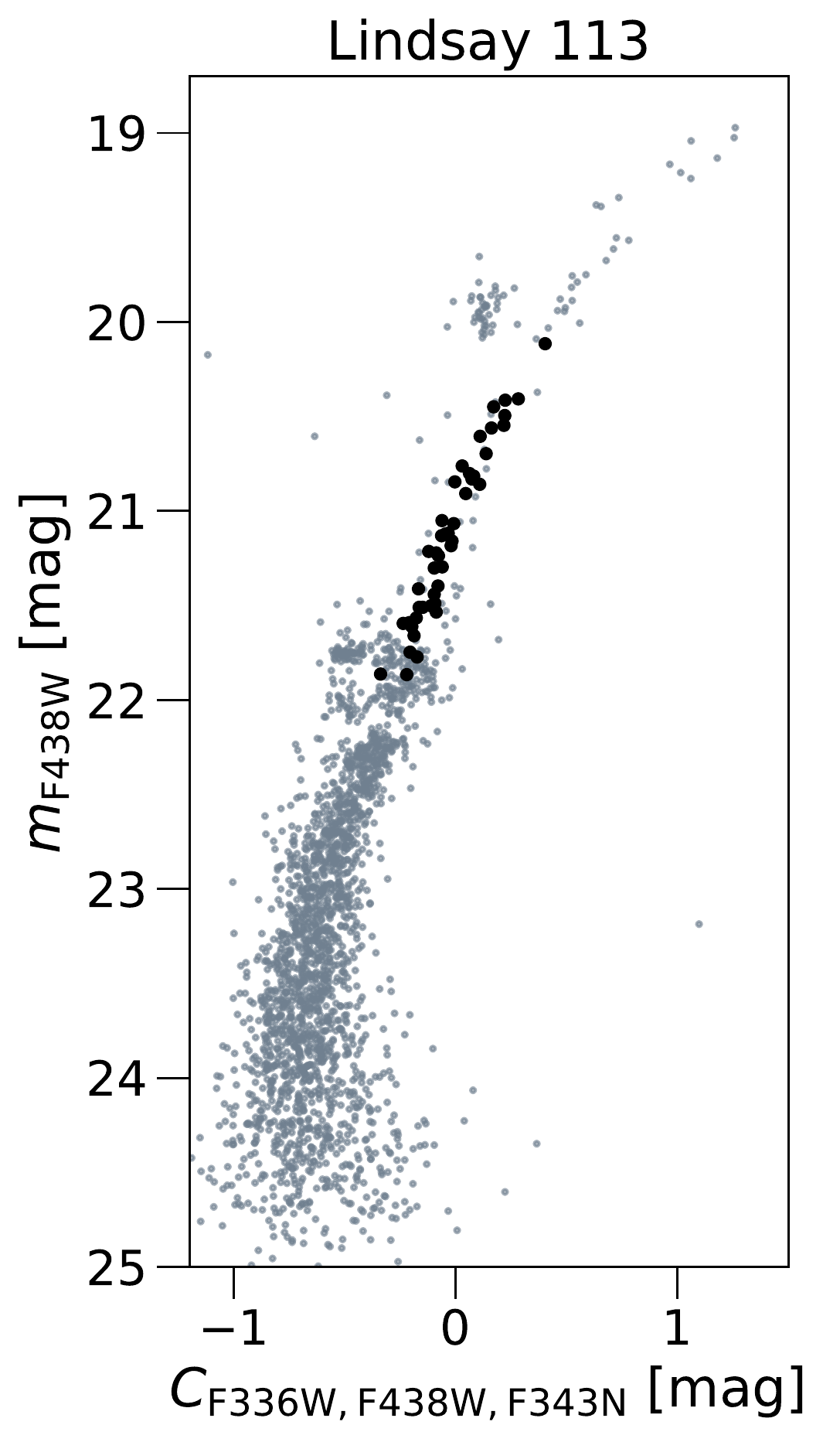}
\includegraphics[scale=0.38]{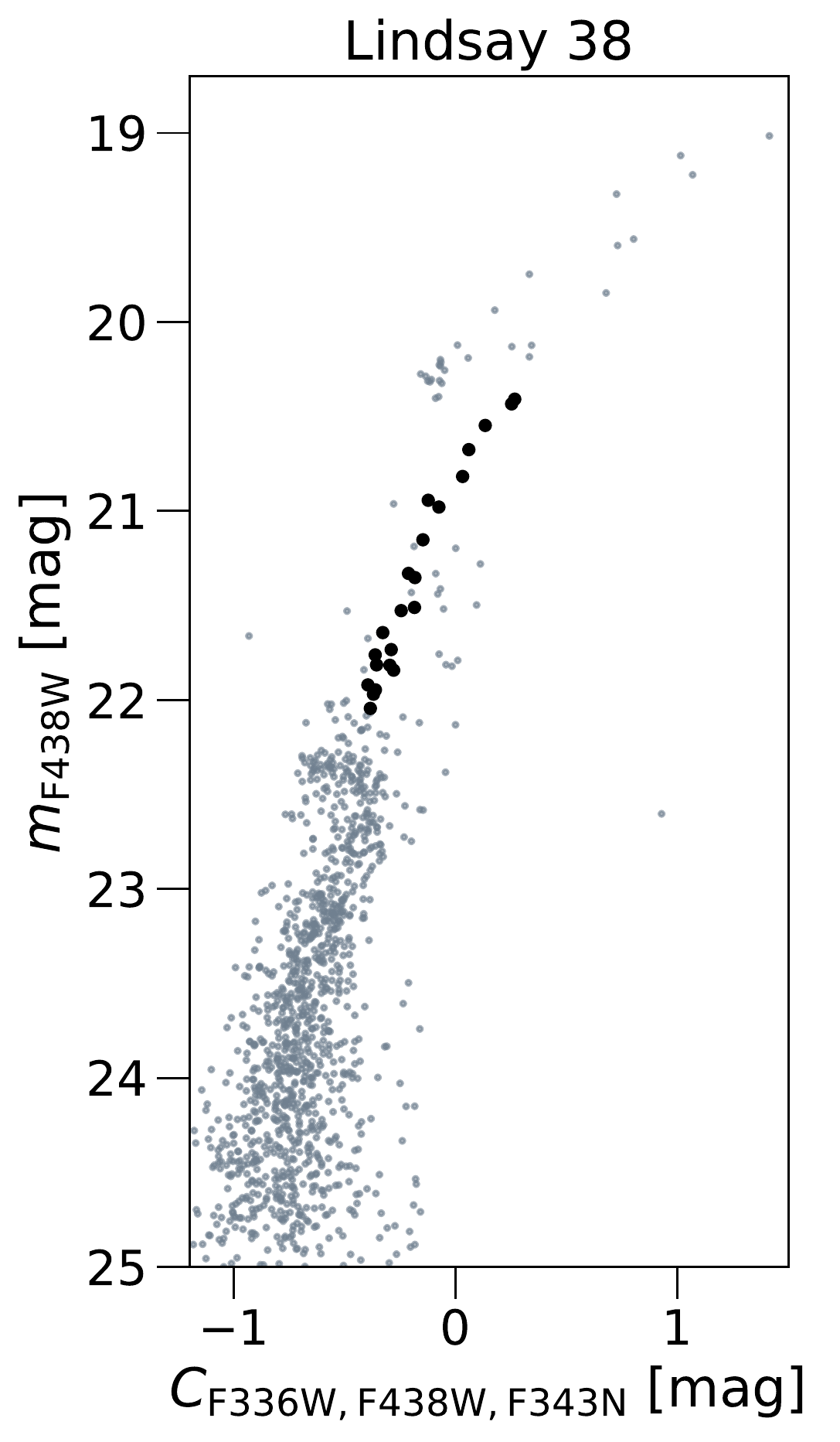}
\caption{\cubun vs. \b CMDs for the clusters analysed in this paper. Black circles indicate the final selected RGB stars.} 
\label{fig:cubun_RGB}
\end{figure*}

%No evident split can be observed in the UV CMDs of all the clusters. 
We used the pseudo-colour \cubun to look for a broadening in the RGB, and to make a homogeneous comparison with the other clusters in our sample (see Papers I to IV). This colour is defined as \cubun$=$(F336W$-$F438W)$-$(F438W$-$F343N) and it has already been proven to be very effective at separating populations with different N abundances (see Papers I and II).  Figure \ref{fig:cubun_RGB} shows the \cubun vs. \b CMDs for all the clusters analysed in this paper, where black filled circles represent the final selected RGB stars in each panel. At a first look, the UV CMDs reveal no clear evidence for splits in the RGB and the RGB looks quite narrow in all cases, except for NGC 2121. To quantify the broadening of the RGB, we took advantage of the AS experiments (\S \ref{subsec:AS}).

\begin{figure*}
\centering
\includegraphics[scale=0.5]{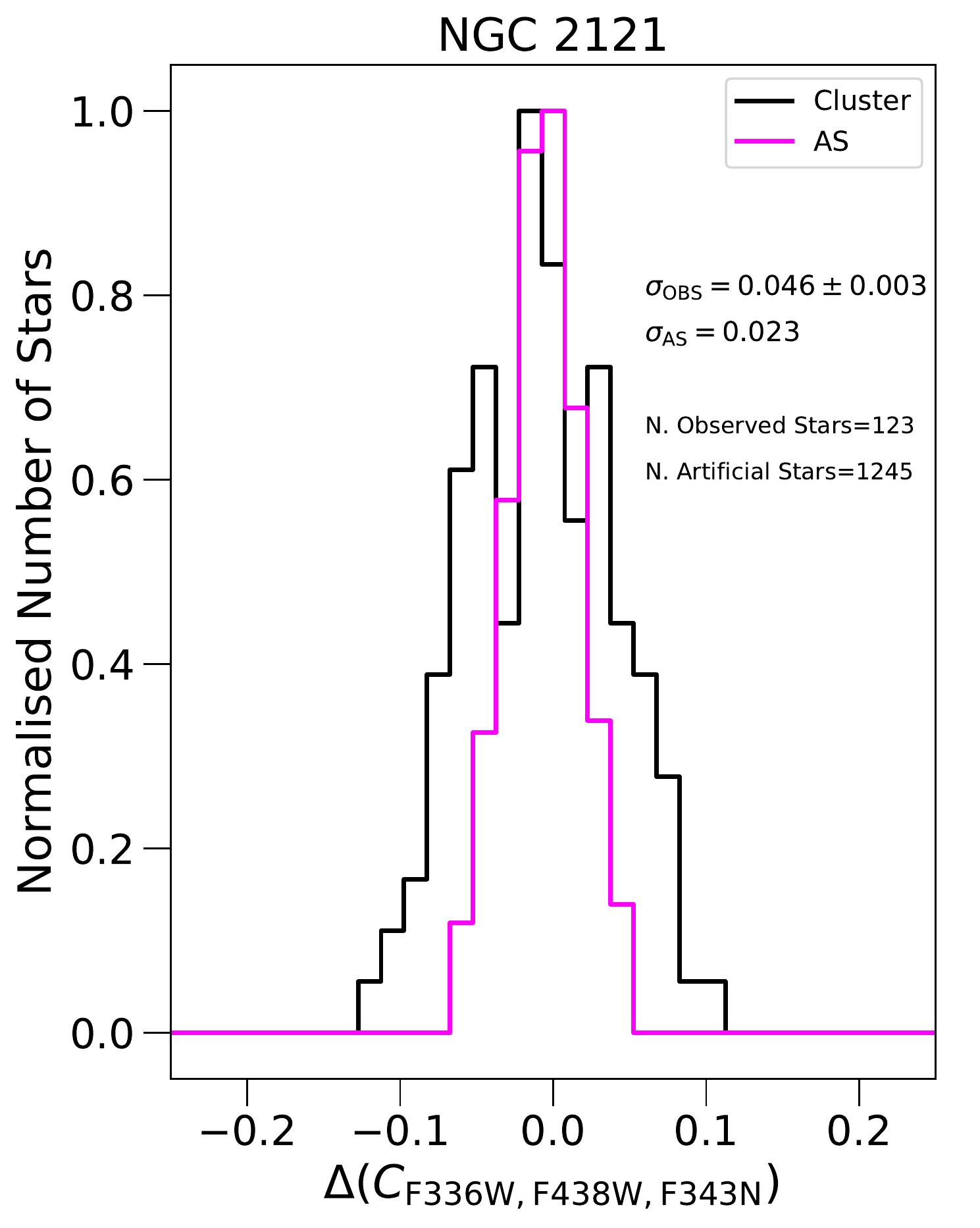}
\includegraphics[scale=0.5]{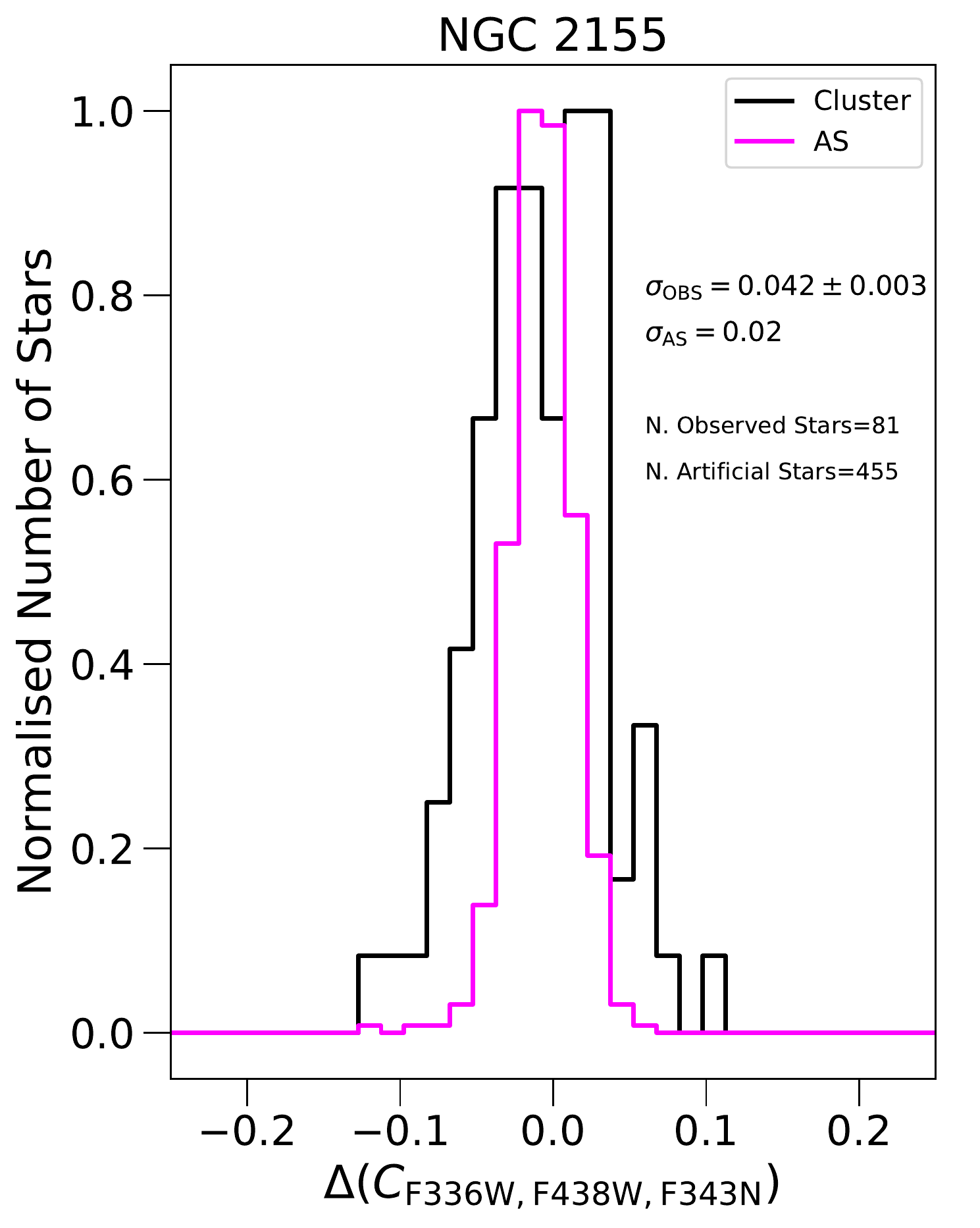}
\includegraphics[scale=0.5]{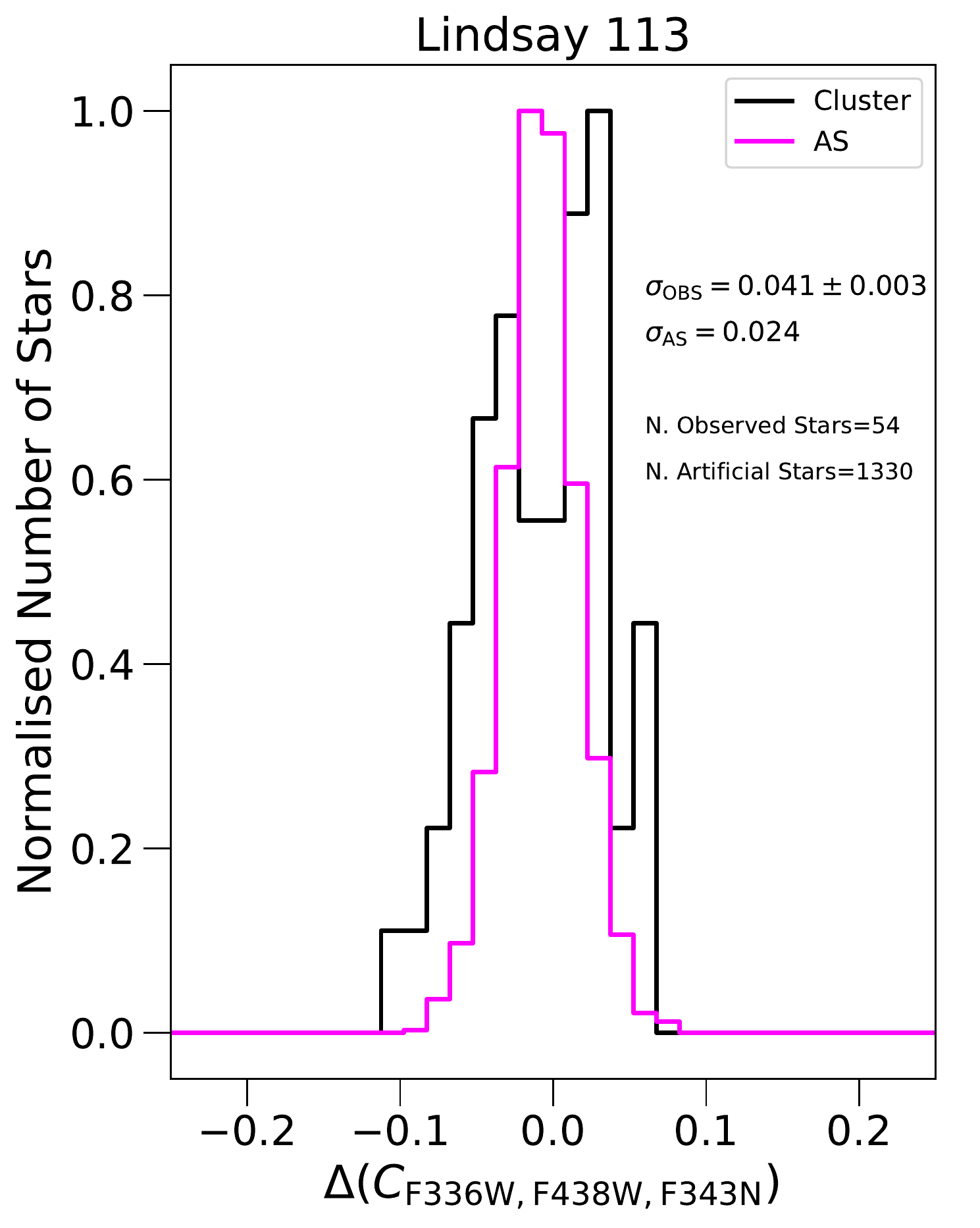}
\includegraphics[scale=0.5]{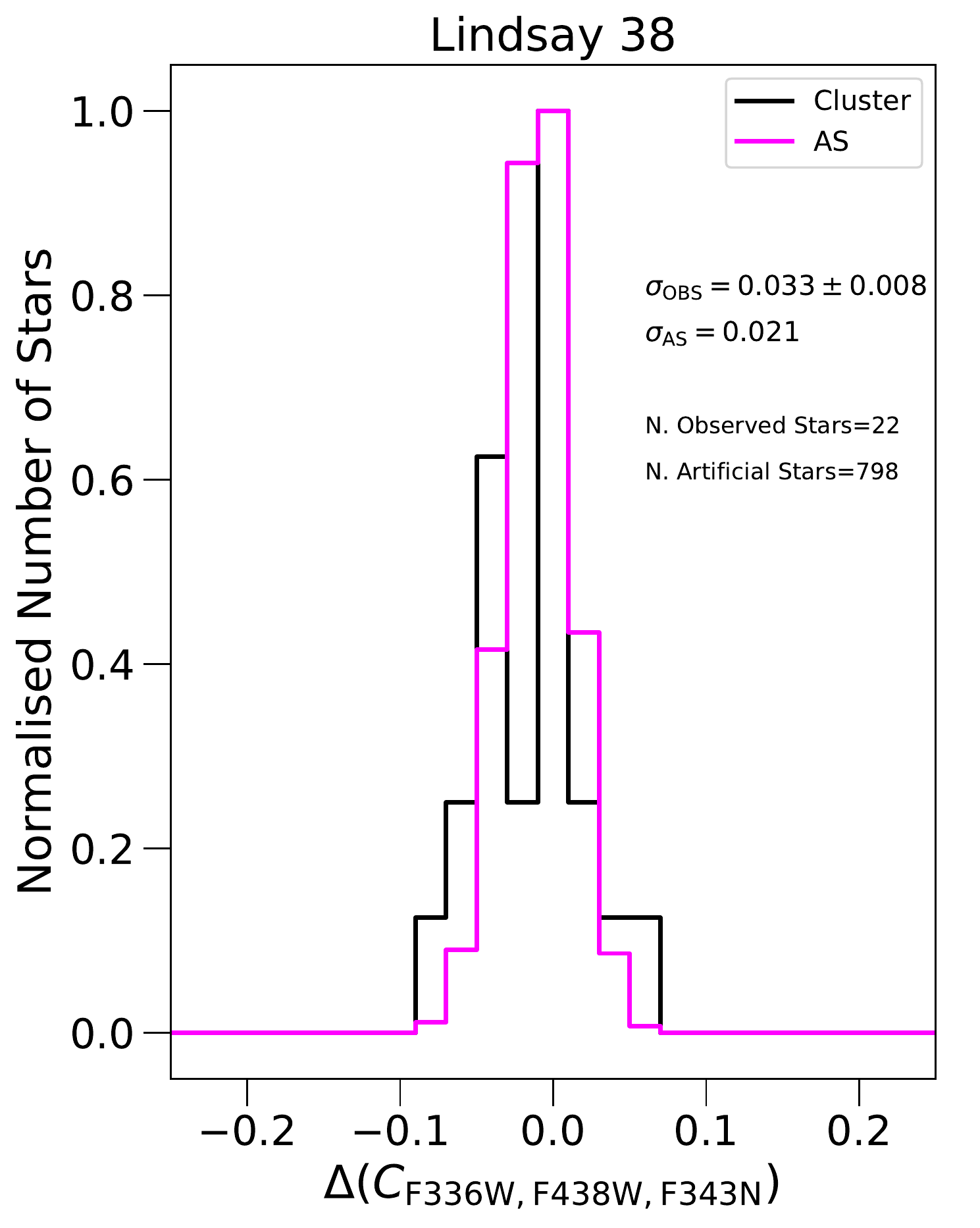}
\caption{Histograms of the distributions of observed (black) and simulated (pink) RGB stars in verticalised \cubun colours for the clusters analysed in this paper. The same bin size was used for real data and AS. Superimposed on the plots are the values of the standard deviations of the distributions. See text for more details.} 
\label{fig:as}
\end{figure*}

We selected RGB stars in the \cubun vs. \b CMDs of the simulated AS catalogues in the same range of magnitude and colours as for the selected observed RGB stars. We then used fiducial lines to verticalise the RGB and obtain $\Delta$(colours) for both observed and simulated catalogues. Figure \ref{fig:as} shows the $\Delta$(\cubun) distributions for observed (black) and simulated (pink) RGB stars for the clusters analysed in this paper. We also calculated the standard deviations of each distribution and these are also superimposed in each panel of Fig. \ref{fig:as}. Errors on standard deviations of the observed distributions were calculated with a bootstrap technique based on 5,000 realizations. 

In all clusters, except for Lindsay 38, a significant broadening in the observed distributions is present when comparing them to the simulated distributions. The observed standard deviations are at least twice as large as the standard deviations of the simulated single stellar population from the AS tests (see $\sigma$s reported in Fig. \ref{fig:as}). 

The case of Lindsay 38 is dominated by poor statistics, as the RGB is composed by $\sim$20 stars. By looking at Fig. \ref{fig:as}, there are no signs of evident broadening in the RGB of Lindsay 38. The standard deviation of the observed distribution is comparable with what we expect from the simulated AS distribution, within the errors. 

Hence, the fact that the distributions of NGC 2121, NGC 2155 and Lindsay 113 are broader than what is expected from a single stellar population, suggests that N variations are present in the RGB stars of such clusters. Based on the current dataset and error estimation, no N variations are instead found in Lindsay 38.  A comparison with the other clusters in
our HST survey will be made in the next Section (\S \ref{sec:results}).

We then fit the discrete $\Delta$(\cubun) data with Gaussian Mixture Models (GMM) to identify the presence of multiple Gaussian components in the colour distribution. We thus derived the probability that a bimodal distribution is rejected for each cluster. Within our observational uncertainties, we find p-values larger than 25\% for all clusters, which demonstrates that bimodality is unlikely in all cases. This was obtained with a parametric bootstrap technique by using the GMM code by \cite{gnedin10}.

When comparing observations to AS catalogues we should note that the errors obtained from AS experiments are systematically underestimated. The main reason is that all AS experiments are simplified to some extent and they are not able to account for all the instrumental sources of noise. The main factor responsible for the error underestimation is likely that the PSF used to fit the artificial stars is also the one used to create them, at odds with what happens with the real stars.
The typical difference between errors from AS and true observational uncertainties has been estimated in previous studies and is of the order of $30-40$\% (see Fig. 4 of \citealt{dalessandro11} and related text and Fig. 21 of \citealt{milone12}). In all clusters, except for Lindsay 38, we observe that the width of the observed distributions is 
$\gtrsim$50\% of the width of the AS distributions, thus we can safely say that a broadening (which is not due to photometric errors) is present in NGC 2121, NGC 2155 and Lindsay 113. Our results for NGC 2121 agree with the conclusions of \cite{li19}. 

\subsection{Age Determination}
\label{subsec:age}

To estimate the age of the clusters in our sample, we superimposed BaSTI isochrones (``A Bag of Stellar Tracks and Isochrones", \citealt{pietrinferni04}) on the optical CMDs of the clusters analysed in this paper. 
We decided to assume average and fixed distance moduli for the LMC and SMC to minimize the number of free parameters involved in the age determination. We assume $(m-M)_{\rm LMC}=18.477$ \citep{lmcdm} and 
$(m-M)_{\rm SMC}=18.965$ \citep{smcdm}. 

\begin{figure*}
\centering
\includegraphics[scale=0.38]{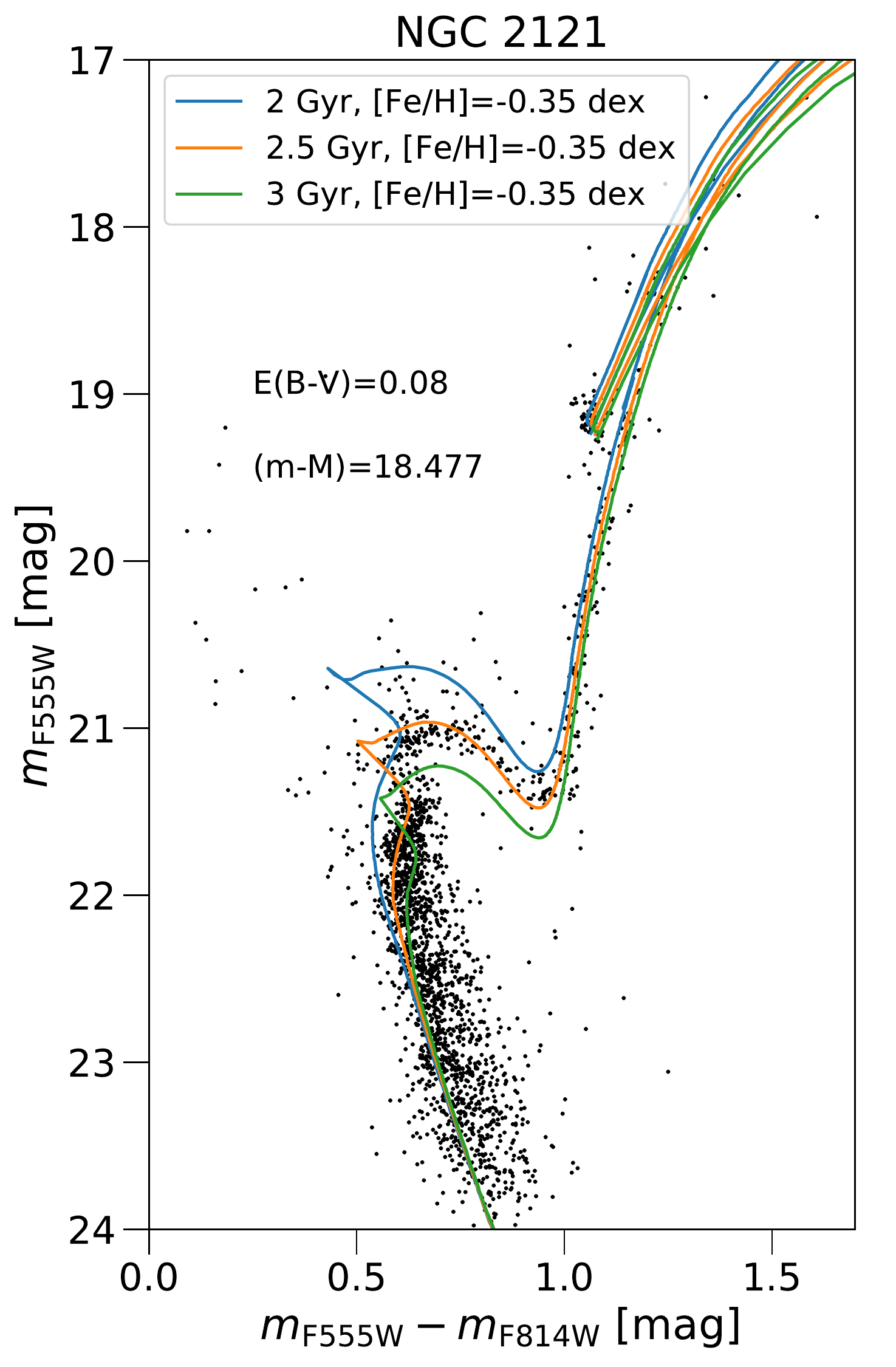}
\includegraphics[scale=0.38]{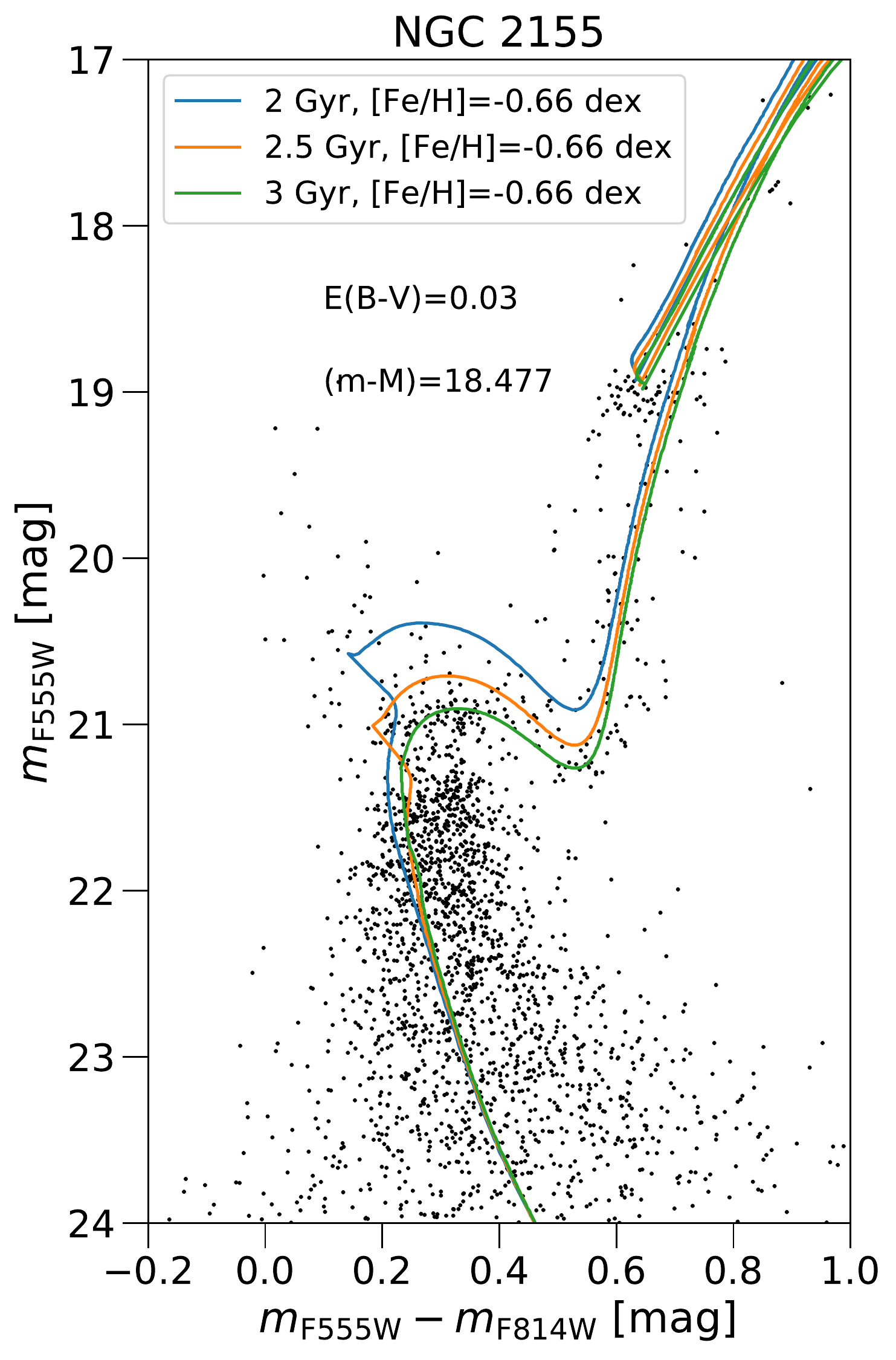}
\includegraphics[scale=0.38]{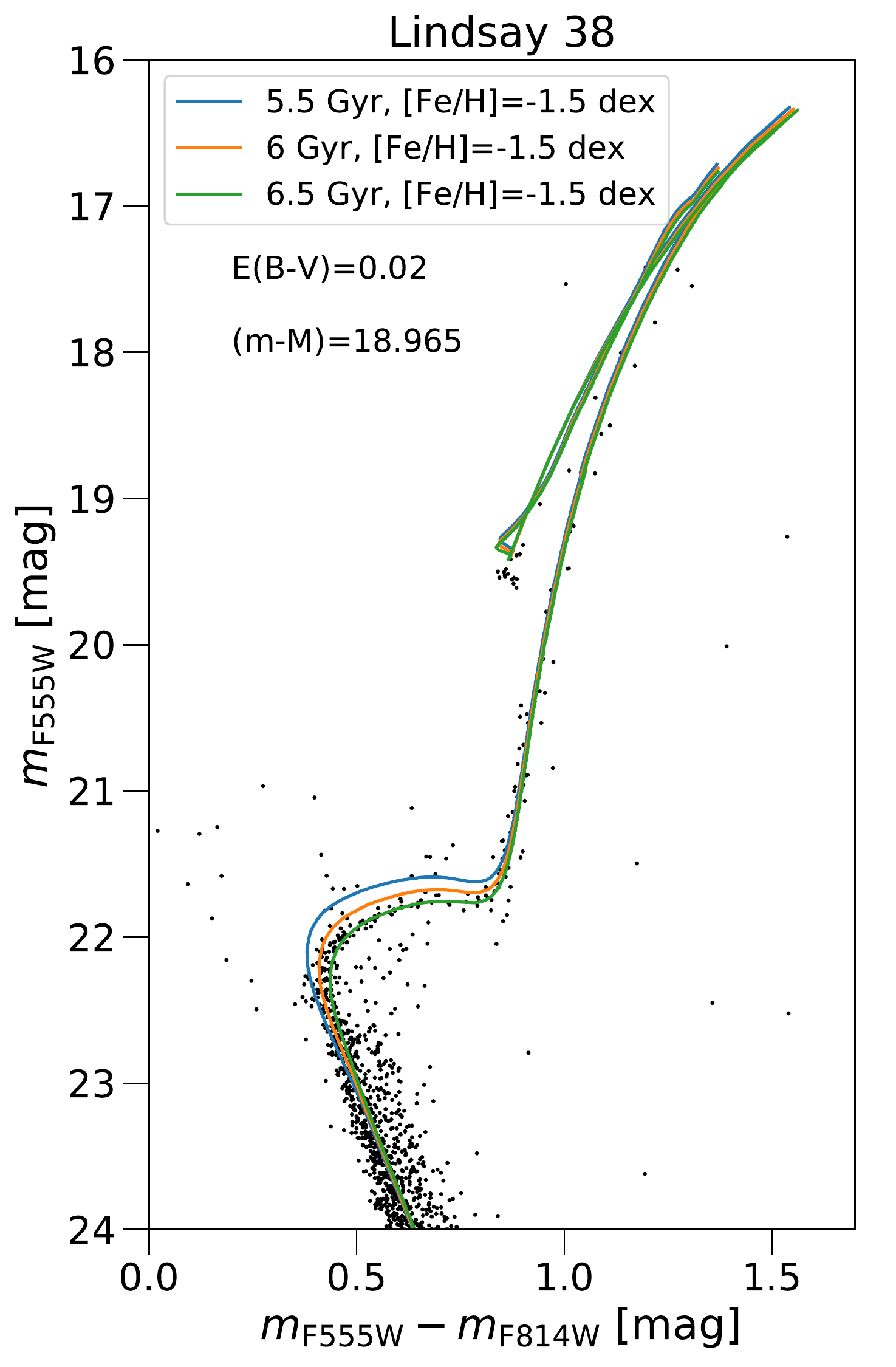}
\caption{\vi vs. \v CMDs for NGC 2121, NGC 2155 and Lindsay 38, respectively from left to right. The blue, orange and green curves represent BaSTI isochrones at different ages (see legend). The metallicity adopted for each cluster is reported in the legend. Finally, the values of the extinction $E(B-V)$ and distance modulus $(m-M)$ are shown in the upper left part of the plots.} 
\label{fig:age}
\end{figure*}

Several isochrones with different metallicities have been used for the fitting of each cluster. The metallicity was chosen to best match simultaneously the RGB and MS.
Figure \ref{fig:age} shows the \vi vs. \v \, CMDs for NGC 2121 and Lindsay 38 and the $m_{\rm F450W}-m_{\rm F555W}$ vs.
$m_{\rm F555W}$ CMD for NGC 2155. Superimposed on the data are three isochrones at different ages, where certain values of metallicity [Fe/H] and extinction $E(B-V)$ were adopted. 
For NGC 2121, we found that the best fit parameters reproducing the shape of the CMD in all its evolutionary stages are: age $\simeq 2.5$ Gyr, metallicity [Fe/H] $= -0.35$ dex, and extinction value $E(B-V)=0.08$ mag. 
We find a similar best fit age for NGC 2155 ($\sim 2.5$ Gyr) along with a metallicity of [Fe/H] $= -0.66$ dex and $E(B-V)=0.03$ mag. For these clusters, we used BaSTI isochrones that account for the effects of core convective overshooting during the central H-burning stage.

For Lindsay 38, the best fit parameters we found by fitting canonical BaSTI isochrones are the following: age $\simeq$ 6.5 Gyr, [Fe/H] $= -1.5$ dex, $E(B-V)=0.02$ mag. 

\begin{figure*}
\centering
\includegraphics[scale=0.42]{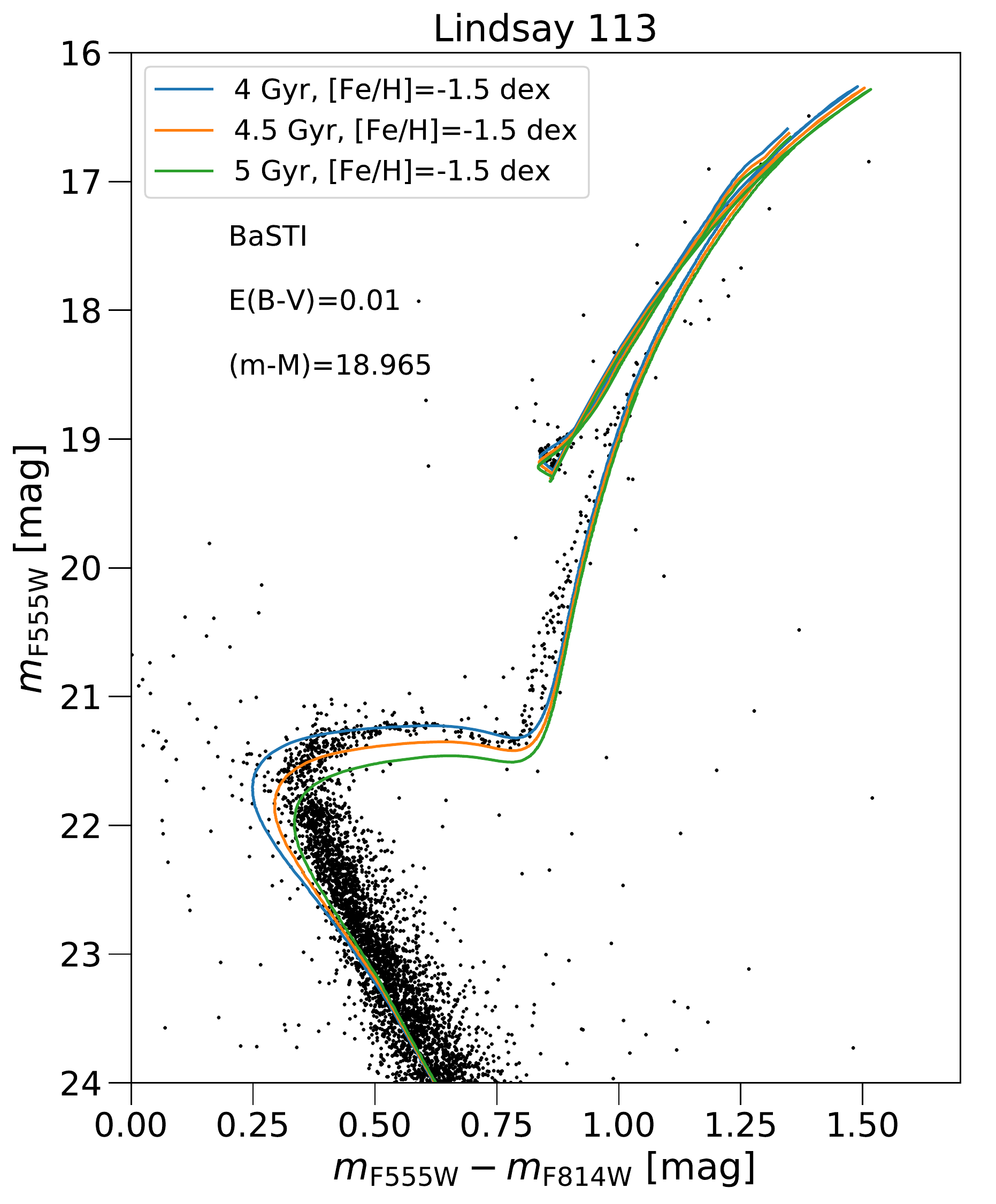}
\includegraphics[scale=0.42]{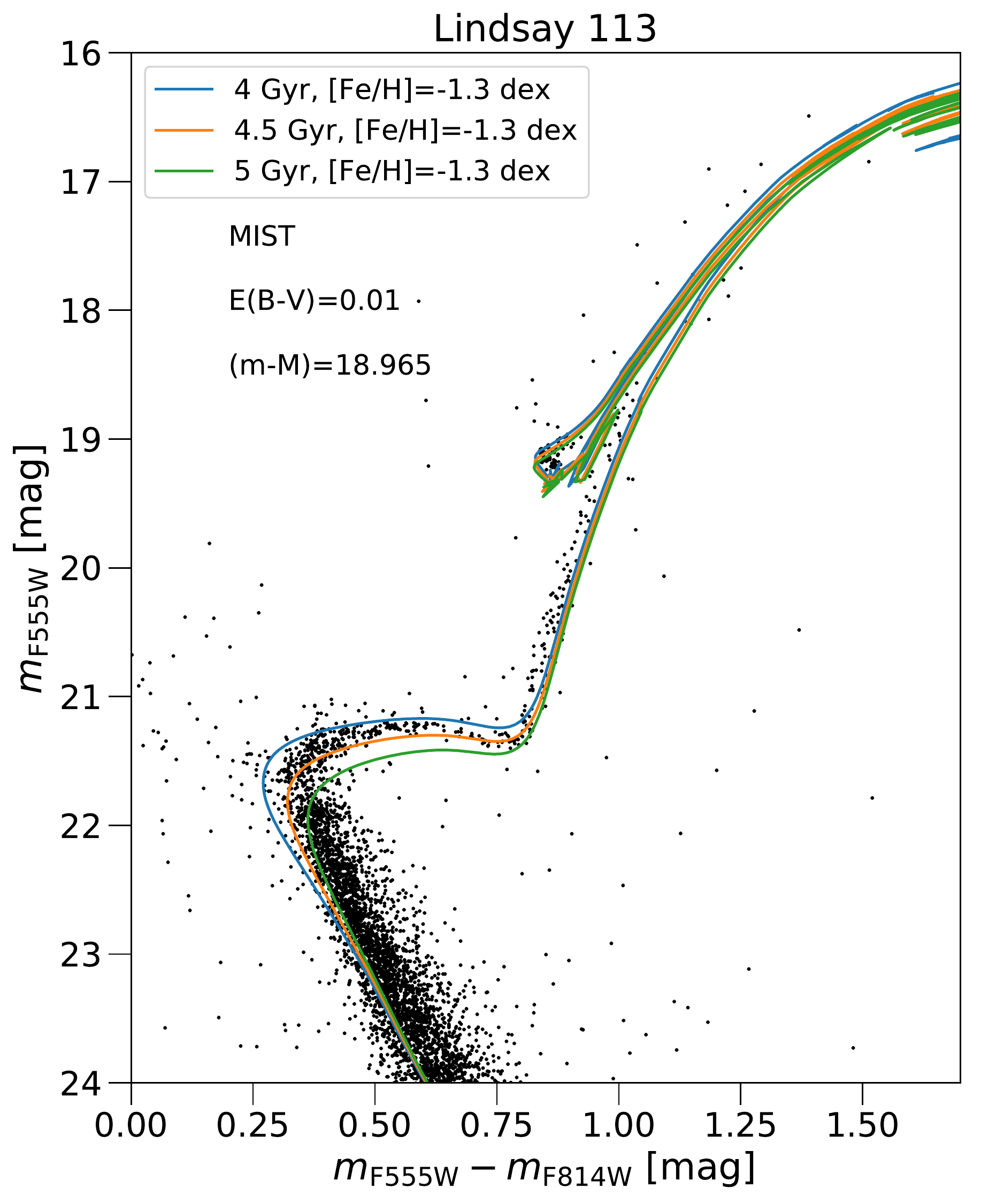}
\caption{\vi vs. \v CMDs for Lindsay 113. The blue, orange and green curves represent BaSTI (left) or MIST (right) isochrones at different ages. The adopted metallicity is reported in the legend for each panel. Finally, the values of the extinction $E(B-V)$ and distance modulus $(m-M)$ are shown in the upper left part of the plots.} 
\label{fig:agel113}
\end{figure*}

BaSTI isochrones on Lindsay 113 could not fit well both MS, RGB and the horizontal branch (HB) at the same time, thus we also explored MIST isochrones (``Mesa Isochrones and Stellar Tracks", \citealt{dotter16, choi16}). 
Figure \ref{fig:agel113} shows the \vi vs. \v \, CMDs for Lindsay 113, where BaSTI (left) and MIST (right) isochrones at different ages are superimposed. 
By adopting the same extinction, we find that there is no considerable difference between the results we get either with BaSTI or MIST. Thus, we found that the best isochrones reproducing the CMD are the 4-4.5 Gyr MIST isochrones with [Fe/H] $= -1.3$ dex and $E(B-V)=0.01$ mag (respectively blue and orange curves in Fig. \ref{fig:agel113}). 
We also note that for NGC 2155 and Lindsay 38, the HB is not matched perfectly. 
A better fit could be reached by slightly changing the cluster distance moduli. However, we conservatively decided to keep them fixed as the required changes have only a small impact on the derived ages.

Table \ref{tab:info} provides information about the parameters adopted for the clusters analysed in this paper. Values of cluster masses from the literature are also reported. 

\begin{table*}
\caption{Adopted values of age, metallicity, distance modulus and reddening for the clusters analysed in this paper. The reported cluster masses are taken from the literature.}
\label{tab:info}
\centering          
    \begin{tabular}{c c c c c c c}    
        \toprule\toprule
        Cluster Name  & Age & [Fe/H] & $(m-M)$ & $E(B-V)$ & Mass & Mass Ref. \\
                             & (Gyr) & (dex) & (mag) & (mag) & ($\times 10^5$ \msun) &  \\
        \midrule
        NGC 2121  & 2.5 & -0.35 & 18.477$^*$ & 0.08 & 1 & (1)\\
        NGC 2155  & 2.5 & -0.66 & 18.477$^*$ & 0.03 & 0.36 & (1)\\
        Lindsay 113  & 4.5 & -1.3 & 18.965$^\dagger$ & 0.01 & 0.23 & (2)\\
        Lindsay 38 & 6.5 & -1.5 & 18.965$^\dagger$ & 0.02 & 0.15 & (3)\\
       \bottomrule\bottomrule
        \end{tabular}
         \\
        (1)~\citet{mclaughlin05}; 
        (2)~\citet{chantereau19};
        (3)~\citet{glatt11}.
        $^*$: fixed value from \cite{lmcdm} (LMC) 
        $^\dagger$: fixed value from \cite{smcdm} (SMC).
\end{table*}

The results shown here are fairly consistent with the literature. 
\cite{glatt08} report an age of $6.5\pm0.5$ Gyr and a metallicity [Fe/H]=-1.5 dex for Lindsay 38 by using the Dartmouth isochrones. Also, \cite{rich01} report an age of 
$3.2\pm0.5$ Gyr for both NGC 2121 and NGC 2155 by using the Padova isochrones, slightly older than what we found, assuming [Fe/H]=-0.68 dex and using Girardi isochrones. Finally, \cite{mighell98} report an age of $4-5$ Gyr for Lindsay 113, with a [Fe/H]$=$-1.2 dex.

From spectroscopy of red giant stars, \cite{grocholski06} find a [Fe/H]=-0.5 dex for NGC 2121 and NGC 2155. We find that NGC 2121 is slighlty more metal rich ([Fe/H]=-0.35 dex, see Table \ref{tab:info}), although errors due to the employment of different methods and isochrones need to be taken into account. Slightly different values have been reported in the literature for Lindsay 113, from [Fe/H]=-1.2 dex \citep{dacostaandh98} to [Fe/H]=-1.03 dex \citep{parisi15}, which are also fairly consistent to what we find ([Fe/H]=-1.3 dex, see Table \ref{tab:info}). Finally, no spectroscopic estimates for the metallicity of Lindsay 38 is reported in the literature so far, to the best of our knowledge.

\section{Results}
\label{sec:results}

We combined the results obtained for the clusters analysed in this paper (namely NGC 2121, NGC 2155, Lindsay 113 and Lindsay 38) with those obtained in Papers I to IV (i.e. NGC 419, 1783, 1806, 1846, 1978, 416, 339, 121, Lindsay 1). 
Finally, we added three ancient clusters ($\gtrsim 12$ Gyr) located in the MW, namely NGC 2419, M15 and 47 Tuc. 

\begin{figure*}
\centering
\includegraphics[scale=0.31]{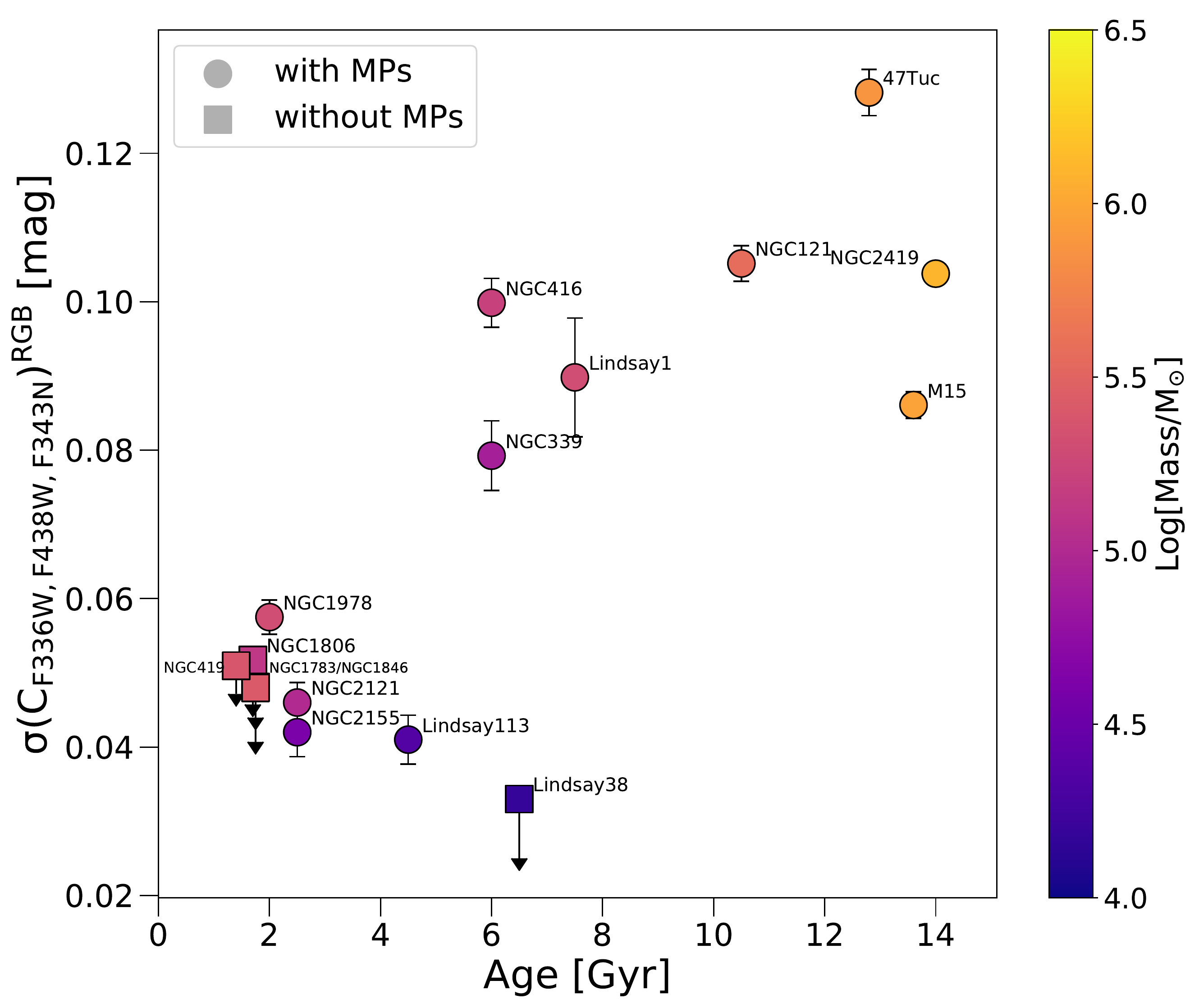}
\includegraphics[scale=0.31]{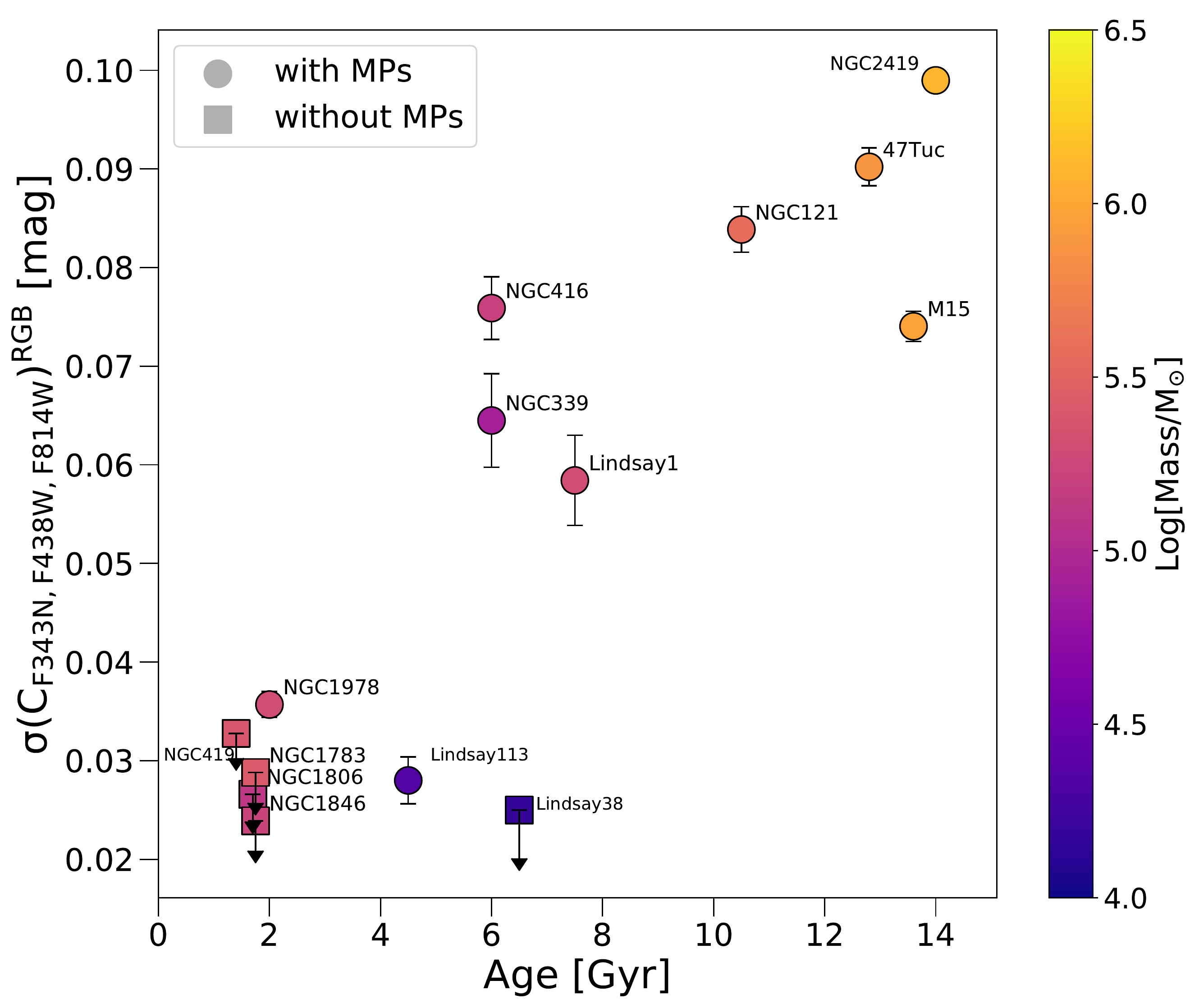}
\caption{Standard deviation of the RGB observed distribution in \cubun (left) and \cunbi (right) colours as a function of cluster age for all the clusters in our HST survey plus 47 Tuc, NGC 2419 and M15. Circles represent cluster with MPs, while squares indicate clusters with no MPs. Data are colour-coded by cluster mass.} 
\label{fig:age1}
\end{figure*}

We calculated the standard deviation of the verticalised distribution of bona-fide RGB stars selected as described in Section \ref{sec:analysis} in \cubun colours for the entire sample. The left panel of Figure \ref{fig:age1} reports the standard deviation as a function of cluster age. Circles indicate clusters with MPs, while squares represent clusters with no significant detection of MPs. Data are colour-coded by cluster mass. Errors on standard deviations were calculated with a bootstrap technique based on 5,000 realizations. 

It is interesting to observe that older clusters show much wider RGBs with respect to the younger ones, representative of more extreme populations. We find that the standard deviations in \cubun\, of the sample analysed in this paper (namely NGC 2121, NGC 2155, Lindsay 113 and Lindsay 38) are comparable, within the errors, with the standard deviations of the clusters that are aged $\lesssim 2$ Gyr. Note that to establish whether chemical anomalies
are present in the clusters of our sample, we compare the RGB width ($\sigma$) with expectations from photometric errors.
However, it is also likely that clusters younger than 2 Gyr might potentially hide smaller N variations that are not detectable by current photometric studies.
Future spectroscopic observations or higher precision photometry will be crucial to understand if this may be the case or not.

Furthermore, it appears that there is not a continuous trend between $\sigma$ and cluster age, as clusters in the age range 2.5$-$4 Gyr have narrower widths of the RGB with respect to NGC 1978, for instance. Nonetheless, other parameters need to be taken into account. 

Cluster mass (at the present day) has been already established to play a fundamental role in the chemical anomalies picture (e.g., \citealt{milone17}), with the extension of the abundance variations becoming larger with increasing stellar mass. %Masses are taken from \cite{gnedin97}, \cite{mclaughlin05}, \cite{goudfrooij14}, \cite{krause16}.
Masses for the MCs clusters are taken from \cite{gnedin97}, \cite{mclaughlin05}, \cite{goudfrooij14}, \cite{krause16}, while the masses for the galactic GCs are from \cite{bh18}.
We note that the cluster mass of our sample is no longer relatively constant. The galactic GCs are 5-10 times more massive than our previous sample, while the new MCs sample reported in this paper is lower mass, by factors of 2-5.

Lindsay 38 is old enough ($\sim$6 Gyr) that one would expect a broader RGB, if age would be the only parameter correlated to abundance variations.
However, this cluster also has a lower mass compared to NGC 339 or NGC 416, by almost one order of magnitude. %We will further discuss our results in Section \S \ref{sec:disc}.

We also explored the behaviour of the RGB by using the pseudo colour \cunbi$\equiv (F343N-F438W)-(F438W-F814W)$ which was used in our previous HST survey study (Papers III and IV). The right panel of Figure \ref{fig:age1} reports the standard deviation of the RGB distributions in verticalised \cunbi colours as a function of cluster age, colour-coded by cluster mass.
However, we did not include two clusters of our sample in this plot, namely NGC 2121 and NGC 2155. For the former, we analysed the errors in the WFPC2 F555W and F814W filters and we noticed that these are more than twice as much compared to the ACS optical filter errors for Lindsay 38 and Lindsay 113. For this reason, we decided not to show the results for NGC 2121 in the \cunbi\, plots. Regarding NGC 2155, we do not have the necessary filters, as only WFPC2 observations in F450W and F555W bands are available (see Table \ref{tab:log}). New optical observations are clearly needed to fully characterise those two clusters. 
The $\sigma$(\cunbi) vs. age plot is consistent with what we found in \cubun\, colours. This contributes to strengthen the result that a correlation between N spread and cluster age is present.

In Cabrera-Ziri et al. (in preparation) we will present a detailed modelling of the effect of age (i.e., the effective temperature of the RGB) and metallicity on the measured widths of \cubun and \cunbi .  However, for the purposes of this paper, the models confirm that \cunbi is essentially independent of age and also of [Fe/H] down to the regime of metal-poor Galactic GCs. Any observed variation/relationship between \cunbi and cluster age can then be attributed to a signature of N enrichment.  There is a small effect of [Fe/H] on \cubun , in the sense that lower [Fe/H] values result in smaller $\Delta$(\cubun) values, but as the two Galactic GCs M15 and NGC 2419 have much lower [Fe/H] but larger $\sigma$(\cubun) values, it is clear that N variations are the driver with age in that diagram as well.

\section{Discussion and conclusions}
\label{sec:disc}

The origin of the unusual chemical patterns typically found in GC stars has remained an unsolved puzzle so far. 
Although much effort has been put into developing new scenarios (e.g. \citealt{gieles18,breen18,howard18}), no consensus has been reached and many observational results remain unexplained (see \citealt{aarev18}).
The exploration of whether a star cluster hosts MPs based on certain cluster properties has been an important avenue of investigation. It is now established that (present day) cluster mass is a fundamental property controlling the extent of which MPs are present, with the star-to-star abundance variations becoming more severe with increasing cluster mass (e.g., \citealt{bragaglia12,schiavon13,milone17}). On the other hand, we also
know that mass cannot be the only parameter which comes into play, as many massive star clusters, although much younger than ancient GCs, 
do not show evidence for the chemical anomalies \citep{mucciarelli08,cabrera16,lardo17,martocchia17,martocchia18a}. 

To shed light onto this, we planned a photometric survey to target star clusters that are as massive as old GCs, but significantly younger. %, i.e. they span from $\sim$ 1.5 up to 11 Gyr in age. 
In this paper, we reported on the photometric analysis of new HST UV images for four clusters in the MCs, namely NGC 2121, NGC 2155, Lindsay 113 and Lindsay 38, %obtained as part of our UV survey of MPs in Magellanic Cloud clusters. NGC 2121 and NGC 2155 are located in the LMC, while Lindsay 113 and Lindsay 38 are located in the SMC.
These clusters have a mass a few times $10^4$\msun \, except for NGC 2121 which is $\sim 10^5$\msun (see Table \ref{tab:info}) and they are aged between $\sim$2.5 and $\sim$6 Gyr.

The UV CMDs of each cluster (see Fig. \ref{fig:cubun_RGB} for the \cubun vs. \b CMDs) reveal no presence of splits in the RGBs. We quantified the broadening of the RGB by comparing the observed verticalised distributions of RGB stars with artificial RGB stars (\S \ref{sec:analysis}). Three out of four clusters in the sample show a significant broadening with respect to photometric errors in \cubun colours, i.e. colours that are sensitive to N variations; Lindsay 38 is the only cluster of the sample whose RGB width is compatible with the errors.
Thus, we add three intermediate-age clusters to our HST survey that show MPs in the form of N spread, namely Lindsay 113, NGC 2121 and NGC 2155.

In Paper IV we found a correlation between cluster age and N enhancement as inferred from photometry, for 9 clusters in the sample. Here we expand our sample to 16 clusters by adding also three GCs (age $\gtrsim$ 12 Gyr) from the MW, namely NGC 2419, M15 and 47 Tuc. We calculate the standard deviation of the verticalised RGB distribution in \cubun and \cunbi colours and we plot this quantity as a function of cluster age. Our results are shown in Fig. \ref{fig:age1}. We find that older clusters show larger widths of the RGB, thus larger N enhancement. 
%Therefore, it appears that older clusters show more extreme populations than the younger clusters, confirming the results we previously found. 
The addition of 7 clusters to the previous sample strengthens the idea that cluster age plays a role in shaping the properties of MPs in GCs. 

However, the exact role of age is currently unknown. It could be suggested that the onset of MPs is due to an evolutionary effect. In our sample we are comparing stars of different masses along the RGB. Some unidentified process operating only in
stars less massive than 1.5 \msun (the mass of a RGB star at
$\sim$2 Gyr) may be responsible for the formation of the chemical
anomalies. %although this would need to act only in dense
%cluster environments as no field stars in the same mass range
%show MPs.
Accordingly, we tentatively argue that chemical anomalies
could be expected to be found in stars with masses below
1.5 \msun\, on the main sequence of young clusters ($<$ 2 Gyr, c.f.
\S 5 point iv, \citealt{cabrera16}).

We note however that our observations are effectively probing N variations among RGB stars within our sample of clusters.  An alternative explanation might  therefore be linked to stellar evolutionary processes where the observed surface abundance of N in RGB stars may be affected.  If this is the case, we may expect to find N-spreads on the main sequence of clusters that do not correspond to the spreads observed along the RGB. This can also be tested by looking at elements less likely to be affected by stellar evolution, e.g., Na or Al.

It is interesting to note that cluster age and cluster mass seem to work simultaneously.
We find that a difference in cluster mass also has an impact at younger ages. As already argued in \S \ref{sec:results}, we observe that Lindsay 38 has a similar age to NGC 339 and NGC 416 but its RGB is less than half as wide. The mass of Lindsay 38 is estimated to be $\sim 10^4$ \msun \citep{glatt11}, an order of magnitude smaller than those of NGC 339 and NGC 416.

However, since all of the clusters in our sample belong to the MCs, it is also possible that the 
appearance of MPs at 2 Gyr could be due to an unknown environmental effect. It would be extremely interesting to test the presence of MPs in clusters beyond the MW and its satellites, but this remains difficult, and new techniques based on integrated light will likely be necessary.

%For instance, clusters younger than 2 Gyr might potentially hide smaller N variations that are not detectable by current photometric studies.
%However, future observations will be crucial to understand if this may be the case or not.

%The results of this ongoing survey have shown that the MPs phenomenon is not a cosmological effect and its onset does not depend on special conditions only present in the early Universe. 

\section*{Acknowledgements}

E.D. acknowledges financial support from the Leverhulme Trust Visiting Professorship Programme VP2-2017-030. E.D. is also grateful for the warm hospitality of LJMU where part of this work was performed. 
V.K-P. gratefully acknowledges financial support for this project provided
by NASA through grant HST-GO-14069 for the Space Telescope Science Institute, which is operated by the Association
of Universities for Research in Astronomy, Inc., under NASA contract NAS526555. 
C.L. thanks the Swiss National Science Foundation for supporting this research through the Ambizione grant number PZ00P2\_168065. N.B. gratefully acknowledges financial support from the Royal Society (University Research Fellowship) and the European Research Council
(ERC-CoG-646928-Multi-Pop). C.U. and W.C. gratefully acknowledges financial support from European Research Council (ERC-CoG-646928-Multi-Pop). W. Chantereau acknowledges funding from the Swiss National Science Foundation under grant P400P2\_183846.
F.N. acknowledges support from the European Research Council (ERC) under European Union's Horizon 2020 research and innovation programme (grant agreement No 682115). 
Support for this work was provided by NASA through Hubble Fellowship grant HST-HF2-51387.001-A awarded by the Space Telescope Science Institute, which is operated by the Association of Universities for Research in Astronomy, Inc., for NASA, under contract NAS5-26555. Finally, we thank the referee for the constructive feedback and suggestions that helped us improve the manuscript.

%%%%%%%%%%%%%%%%%%%%%%%%%%%%%%%%%%%%%%%%%%%%%%%%%%

%%%%%%%%%%%%%%%%%%%% REFERENCES %%%%%%%%%%%%%%%%%%

% The best way to enter references is to use BibTeX:

\bibliographystyle{mnras}
\bibliography{hst_5} % if your bibtex file is called example.bib

\begin{thebibliography}{}
\makeatletter
\relax
\def\mn@urlcharsother{\let\do\@makeother \do\$\do\&\do\#\do\^\do\_\do\%\do\~}
\def\mn@doi{\begingroup\mn@urlcharsother \@ifnextchar [ {\mn@doi@}
  {\mn@doi@[]}}
\def\mn@doi@[#1]#2{\def\@tempa{#1}\ifx\@tempa\@empty \href
  {http://dx.doi.org/#2} {doi:#2}\else \href {http://dx.doi.org/#2} {#1}\fi
  \endgroup}
\def\mn@eprint#1#2{\mn@eprint@#1:#2::\@nil}
\def\mn@eprint@arXiv#1{\href {http://arxiv.org/abs/#1} {{\tt arXiv:#1}}}
\def\mn@eprint@dblp#1{\href {http://dblp.uni-trier.de/rec/bibtex/#1.xml}
  {dblp:#1}}
\def\mn@eprint@#1:#2:#3:#4\@nil{\def\@tempa {#1}\def\@tempb {#2}\def\@tempc
  {#3}\ifx \@tempc \@empty \let \@tempc \@tempb \let \@tempb \@tempa \fi \ifx
  \@tempb \@empty \def\@tempb {arXiv}\fi \@ifundefined
  {mn@eprint@\@tempb}{\@tempb:\@tempc}{\expandafter \expandafter \csname
  mn@eprint@\@tempb\endcsname \expandafter{\@tempc}}}

\bibitem[\protect\citeauthoryear{{Bastian} \& {Lardo}}{{Bastian} \&
  {Lardo}}{2018}]{aarev18}
{Bastian} N.,  {Lardo} C.,  2018, \mn@doi [\araa]
  {10.1146/annurev-astro-081817-051839}, \href
  {http://adsabs.harvard.edu/abs/2018ARA%26A..56...83B} {56, 83}

\bibitem[\protect\citeauthoryear{{Bastian} \& {de Mink}}{{Bastian} \& {de
  Mink}}{2009}]{bastiandemink09}
{Bastian} N.,  {de Mink} S.~E.,  2009, \mn@doi [MNRAS]
  {10.1111/j.1745-3933.2009.00696.x}, \href
  {http://adsabs.harvard.edu/abs/2009MNRAS.398L..11B} {398, L11}

\bibitem[\protect\citeauthoryear{{Bastian} et~al.,}{{Bastian}
  et~al.}{2016}]{bastian16}
{Bastian} N.,  et~al., 2016, \mn@doi [MNRAS] {10.1093/mnrasl/slw067}, \href
  {http://adsabs.harvard.edu/abs/2016MNRAS.460L..20B} {460, L20}

\bibitem[\protect\citeauthoryear{{Bastian}, {Kamann}, {Cabrera-Ziri}, {Georgy},
  {Ekstr{\"o}m}, {Charbonnel}, {de Juan Ovelar}  \& {Usher}}{{Bastian}
  et~al.}{2018}]{bastian18}
{Bastian} N.,  {Kamann} S.,  {Cabrera-Ziri} I.,  {Georgy} C.,  {Ekstr{\"o}m}
  S.,  {Charbonnel} C.,  {de Juan Ovelar} M.,   {Usher} C.,  2018, \mn@doi
  [\mnras] {10.1093/mnras/sty2100}, \href
  {http://adsabs.harvard.edu/abs/2018MNRAS.480.3739B} {480, 3739}

\bibitem[\protect\citeauthoryear{{Baumgardt} \& {Hilker}}{{Baumgardt} \&
  {Hilker}}{2018}]{bh18}
{Baumgardt} H.,  {Hilker} M.,  2018, \mn@doi [\mnras] {10.1093/mnras/sty1057},
  \href {http://adsabs.harvard.edu/abs/2018MNRAS.478.1520B} {478, 1520}

\bibitem[\protect\citeauthoryear{{Bellazzini}, {Fusi Pecci}, {Montegriffo},
  {Messineo}, {Monaco}  \& {Rood}}{{Bellazzini} et~al.}{2002}]{bellazzini02}
{Bellazzini} M.,  {Fusi Pecci} F.,  {Montegriffo} P.,  {Messineo} M.,  {Monaco}
  L.,   {Rood} R.~T.,  2002, \mn@doi [\aj] {10.1086/340082}, \href
  {http://adsabs.harvard.edu/abs/2002AJ....123.2541B} {123, 2541}

\bibitem[\protect\citeauthoryear{{Bragaglia}, {Gratton}, {Carretta}, {D'Orazi},
  {Sneden}  \& {Lucatello}}{{Bragaglia} et~al.}{2012}]{bragaglia12}
{Bragaglia} A.,  {Gratton} R.~G.,  {Carretta} E.,  {D'Orazi} V.,  {Sneden} C.,
   {Lucatello} S.,  2012, \mn@doi [\aap] {10.1051/0004-6361/201220366}, \href
  {http://adsabs.harvard.edu/abs/2012A%26A...548A.122B} {548, A122}

\bibitem[\protect\citeauthoryear{{Breen}}{{Breen}}{2018}]{breen18}
{Breen} P.~G.,  2018, \mn@doi [\mnras] {10.1093/mnrasl/sly169}, \href
  {http://adsabs.harvard.edu/abs/2018MNRAS.481L.110B} {481, L110}

\bibitem[\protect\citeauthoryear{{Cabrera-Ziri}, {Lardo}, {Davies}, {Bastian},
  {Beccari}, {Larsen}  \& {Hernandez}}{{Cabrera-Ziri} et~al.}{2016}]{cabrera16}
{Cabrera-Ziri} I.,  {Lardo} C.,  {Davies} B.,  {Bastian} N.,  {Beccari} G.,
  {Larsen} S.~S.,   {Hernandez} S.,  2016, \mn@doi [\mnras]
  {10.1093/mnras/stw1090}, \href
  {http://adsabs.harvard.edu/abs/2016MNRAS.460.1869C} {460, 1869}

\bibitem[\protect\citeauthoryear{{Carretta}, {Bragaglia}, {Gratton},
  {Recio-Blanco}, {Lucatello}, {D'Orazi}  \& {Cassisi}}{{Carretta}
  et~al.}{2010}]{carretta10}
{Carretta} E.,  {Bragaglia} A.,  {Gratton} R.~G.,  {Recio-Blanco} A.,
  {Lucatello} S.,  {D'Orazi} V.,   {Cassisi} S.,  2010, \mn@doi [\aap]
  {10.1051/0004-6361/200913451}, \href
  {http://adsabs.harvard.edu/abs/2010A%26A...516A..55C} {516, A55}

\bibitem[\protect\citeauthoryear{{Chantereau}, {Salaris}, {Bastian}  \&
  {Martocchia}}{{Chantereau} et~al.}{2019}]{chantereau19}
{Chantereau} W.,  {Salaris} M.,  {Bastian} N.,   {Martocchia} S.,  2019,
  \mn@doi [\mnras] {10.1093/mnras/stz378}, \href
  {http://adsabs.harvard.edu/abs/2019MNRAS.484.5236C} {484, 5236}

\bibitem[\protect\citeauthoryear{{Choi}, {Dotter}, {Conroy}, {Cantiello},
  {Paxton}  \& {Johnson}}{{Choi} et~al.}{2016}]{choi16}
{Choi} J.,  {Dotter} A.,  {Conroy} C.,  {Cantiello} M.,  {Paxton} B.,
  {Johnson} B.~D.,  2016, \mn@doi [ApJ] {10.3847/0004-637X/823/2/102}, \href
  {http://adsabs.harvard.edu/abs/2016ApJ...823..102C} {823, 102}

\bibitem[\protect\citeauthoryear{{D'Antona}, {Di Criscienzo}, {Decressin},
  {Milone}, {Vesperini}  \& {Ventura}}{{D'Antona} et~al.}{2015}]{dantona15}
{D'Antona} F.,  {Di Criscienzo} M.,  {Decressin} T.,  {Milone} A.~P.,
  {Vesperini} E.,   {Ventura} P.,  2015, \mn@doi [\mnras]
  {10.1093/mnras/stv1794}, \href
  {http://adsabs.harvard.edu/abs/2015MNRAS.453.2637D} {453, 2637}

\bibitem[\protect\citeauthoryear{{Da Costa} \& {Hatzidimitriou}}{{Da Costa} \&
  {Hatzidimitriou}}{1998}]{dacostaandh98}
{Da Costa} G.~S.,  {Hatzidimitriou} D.,  1998, \mn@doi [\aj] {10.1086/300340},
  \href {http://adsabs.harvard.edu/abs/1998AJ....115.1934D} {115, 1934}

\bibitem[\protect\citeauthoryear{{Dalessandro}, {Lanzoni}, {Beccari},
  {Sollima}, {Ferraro}  \& {Pasquato}}{{Dalessandro}
  et~al.}{2011}]{dalessandro11}
{Dalessandro} E.,  {Lanzoni} B.,  {Beccari} G.,  {Sollima} A.,  {Ferraro}
  F.~R.,   {Pasquato} M.,  2011, \mn@doi [\apj] {10.1088/0004-637X/743/1/11},
  \href {http://adsabs.harvard.edu/abs/2011ApJ...743...11D} {743, 11}

\bibitem[\protect\citeauthoryear{{Dalessandro} et~al.,}{{Dalessandro}
  et~al.}{2014}]{dalessandro14}
{Dalessandro} E.,  et~al., 2014, \mn@doi [ApJl] {10.1088/2041-8205/791/1/L4},
  \href {http://adsabs.harvard.edu/abs/2014ApJ...791L...4D} {791, L4}

\bibitem[\protect\citeauthoryear{{Dalessandro}, {Ferraro}, {Massari},
  {Lanzoni}, {Miocchi}  \& {Beccari}}{{Dalessandro}
  et~al.}{2015}]{dalessandro15}
{Dalessandro} E.,  {Ferraro} F.~R.,  {Massari} D.,  {Lanzoni} B.,  {Miocchi}
  P.,   {Beccari} G.,  2015, \mn@doi [\apj] {10.1088/0004-637X/810/1/40}, \href
  {http://adsabs.harvard.edu/abs/2015ApJ...810...40D} {810, 40}

\bibitem[\protect\citeauthoryear{{Dalessandro}, {Lapenna}, {Mucciarelli},
  {Origlia}, {Ferraro}  \& {Lanzoni}}{{Dalessandro}
  et~al.}{2016}]{dalessandro16}
{Dalessandro} E.,  {Lapenna} E.,  {Mucciarelli} A.,  {Origlia} L.,  {Ferraro}
  F.~R.,   {Lanzoni} B.,  2016, \mn@doi [ApJ] {10.3847/0004-637X/829/2/77},
  \href {http://adsabs.harvard.edu/abs/2016ApJ...829...77D} {829, 77}

\bibitem[\protect\citeauthoryear{{Dalessandro} et~al.,}{{Dalessandro}
  et~al.}{2018}]{dalessandro18}
{Dalessandro} E.,  et~al., 2018, \mn@doi [\aap] {10.1051/0004-6361/201833650},
  \href {http://adsabs.harvard.edu/abs/2018A%26A...618A.131D} {618, A131}

\bibitem[\protect\citeauthoryear{{Dotter}}{{Dotter}}{2016}]{dotter16}
{Dotter} A.,  2016, \mn@doi [ApJS] {10.3847/0067-0049/222/1/8}, \href
  {http://adsabs.harvard.edu/abs/2016ApJS..222....8D} {222, 8}

\bibitem[\protect\citeauthoryear{{Dotter}, {Milone}, {Conroy}, {Marino}  \&
  {Sarajedini}}{{Dotter} et~al.}{2018}]{dotter18}
{Dotter} A.,  {Milone} A.~P.,  {Conroy} C.,  {Marino} A.~F.,   {Sarajedini} A.,
   2018, \mn@doi [\apjl] {10.3847/2041-8213/aae08f}, \href
  {http://adsabs.harvard.edu/abs/2018ApJ...865L..10D} {865, L10}

\bibitem[\protect\citeauthoryear{{Gieles} et~al.,}{{Gieles}
  et~al.}{2018}]{gieles18}
{Gieles} M.,  et~al., 2018, \mn@doi [\mnras] {10.1093/mnras/sty1059}, \href
  {http://adsabs.harvard.edu/abs/2018MNRAS.478.2461G} {478, 2461}

\bibitem[\protect\citeauthoryear{{Gilligan} et~al.,}{{Gilligan}
  et~al.}{2019}]{gilligan19}
{Gilligan} C.~K.,  et~al., 2019, arXiv e-prints, \href
  {http://adsabs.harvard.edu/abs/2019arXiv190401434G} {}

\bibitem[\protect\citeauthoryear{{Glatt} et~al.,}{{Glatt}
  et~al.}{2008}]{glatt08}
{Glatt} K.,  et~al., 2008, \mn@doi [AJ] {10.1088/0004-6256/136/4/1703}, \href
  {http://adsabs.harvard.edu/abs/2008AJ....136.1703G} {136, 1703}

\bibitem[\protect\citeauthoryear{{Glatt} et~al.,}{{Glatt}
  et~al.}{2011}]{glatt11}
{Glatt} K.,  et~al., 2011, \mn@doi [\aj] {10.1088/0004-6256/142/2/36}, \href
  {http://adsabs.harvard.edu/abs/2011AJ....142...36G} {142, 36}

\bibitem[\protect\citeauthoryear{{Gnedin} \& {Ostriker}}{{Gnedin} \&
  {Ostriker}}{1997}]{gnedin97}
{Gnedin} O.~Y.,  {Ostriker} J.~P.,  1997, \mn@doi [\apj] {10.1086/303441},
  \href {http://adsabs.harvard.edu/abs/1997ApJ...474..223G} {474, 223}

\bibitem[\protect\citeauthoryear{{Goudfrooij} et~al.,}{{Goudfrooij}
  et~al.}{2014}]{goudfrooij14}
{Goudfrooij} P.,  et~al., 2014, \mn@doi [\apj] {10.1088/0004-637X/797/1/35},
  \href {http://adsabs.harvard.edu/abs/2014ApJ...797...35G} {797, 35}

\bibitem[\protect\citeauthoryear{{Graczyk} et~al.,}{{Graczyk}
  et~al.}{2019}]{smcdm}
{Graczyk} D.,  et~al., 2019, \mn@doi [\apj] {10.3847/1538-4357/aafbed}, \href
  {http://adsabs.harvard.edu/abs/2019ApJ...872...85G} {872, 85}

\bibitem[\protect\citeauthoryear{{Gratton}, {Carretta}  \&
  {Bragaglia}}{{Gratton} et~al.}{2012}]{gratton12}
{Gratton} R.~G.,  {Carretta} E.,   {Bragaglia} A.,  2012, \mn@doi [\aapr]
  {10.1007/s00159-012-0050-3}, \href
  {http://adsabs.harvard.edu/abs/2012A%26ARv..20...50G} {20, 50}

\bibitem[\protect\citeauthoryear{{Grocholski}, {Cole}, {Sarajedini}, {Geisler}
  \& {Smith}}{{Grocholski} et~al.}{2006}]{grocholski06}
{Grocholski} A.~J.,  {Cole} A.~A.,  {Sarajedini} A.,  {Geisler} D.,   {Smith}
  V.~V.,  2006, \mn@doi [\aj] {10.1086/507303}, \href
  {http://adsabs.harvard.edu/abs/2006AJ....132.1630G} {132, 1630}

\bibitem[\protect\citeauthoryear{{Hollyhead} et~al.,}{{Hollyhead}
  et~al.}{2019}]{hollyhead19}
{Hollyhead} K.,  et~al., 2019, \mn@doi [\mnras] {10.1093/mnras/stz317}, \href
  {http://adsabs.harvard.edu/abs/2019MNRAS.484.4718H} {484, 4718}

\bibitem[\protect\citeauthoryear{{Howard}, {Pudritz}, {Sills}  \&
  {Harris}}{{Howard} et~al.}{2019}]{howard18}
{Howard} C.~S.,  {Pudritz} R.~E.,  {Sills} A.,   {Harris} W.~E.,  2019, \mn@doi
  [\mnras] {10.1093/mnras/stz924}, \href
  {http://adsabs.harvard.edu/abs/2019MNRAS.486.1146H} {486, 1146}

\bibitem[\protect\citeauthoryear{{Kamann} et~al.,}{{Kamann}
  et~al.}{2018}]{kamann18}
{Kamann} S.,  et~al., 2018, \mn@doi [\mnras] {10.1093/mnras/sty1958}, \href
  {http://adsabs.harvard.edu/abs/2018MNRAS.480.1689K} {480, 1689}

\bibitem[\protect\citeauthoryear{{Krause}, {Charbonnel}, {Bastian}  \&
  {Diehl}}{{Krause} et~al.}{2016}]{krause16}
{Krause} M.~G.~H.,  {Charbonnel} C.,  {Bastian} N.,   {Diehl} R.,  2016,
  \mn@doi [\aap] {10.1051/0004-6361/201526685}, \href
  {http://adsabs.harvard.edu/abs/2016A%26A...587A..53K} {587, A53}

\bibitem[\protect\citeauthoryear{{Lardo}, {Cabrera-Ziri}, {Davies}  \&
  {Bastian}}{{Lardo} et~al.}{2017}]{lardo17}
{Lardo} C.,  {Cabrera-Ziri} I.,  {Davies} B.,   {Bastian} N.,  2017, \mn@doi
  [\mnras] {10.1093/mnras/stx628}, \href
  {http://adsabs.harvard.edu/abs/2017MNRAS.468.2482L} {468, 2482}

\bibitem[\protect\citeauthoryear{{Larsen}, {Brodie}, {Grundahl}  \&
  {Strader}}{{Larsen} et~al.}{2014}]{larsen14}
{Larsen} S.~S.,  {Brodie} J.~P.,  {Grundahl} F.,   {Strader} J.,  2014, \mn@doi
  [ApJ] {10.1088/0004-637X/797/1/15}, \href
  {http://adsabs.harvard.edu/abs/2014ApJ...797...15L} {797, 15}

\bibitem[\protect\citeauthoryear{{Larsen}, {Baumgardt}, {Bastian}, {Hernandez}
  \& {Brodie}}{{Larsen} et~al.}{2019}]{larsen19}
{Larsen} S.~S.,  {Baumgardt} H.,  {Bastian} N.,  {Hernandez} S.,   {Brodie}
  J.~P.,  2019, arXiv e-prints, \href
  {http://adsabs.harvard.edu/abs/2019arXiv190201416L} {}

\bibitem[\protect\citeauthoryear{{Li} \& {de Grijs}}{{Li} \& {de
  Grijs}}{2019}]{li19}
{Li} C.,  {de Grijs} R.,  2019, arXiv e-prints, \href
  {http://adsabs.harvard.edu/abs/2019arXiv190400508L} {}

\bibitem[\protect\citeauthoryear{{Mackey}, {Broby Nielsen}, {Ferguson}  \&
  {Richardson}}{{Mackey} et~al.}{2008}]{mackey08}
{Mackey} A.~D.,  {Broby Nielsen} P.,  {Ferguson} A.~M.~N.,   {Richardson}
  J.~C.,  2008, \mn@doi [ApJL] {10.1086/590343}, \href
  {http://adsabs.harvard.edu/abs/2008ApJ...681L..17M} {681, L17}

\bibitem[\protect\citeauthoryear{{Martocchia} et~al.,}{{Martocchia}
  et~al.}{2017}]{martocchia17}
{Martocchia} S.,  et~al., 2017, \mn@doi [MNRAS] {10.1093/mnras/stx660}, \href
  {http://adsabs.harvard.edu/abs/2017MNRAS.468.3150M} {468, 3150}

\bibitem[\protect\citeauthoryear{{Martocchia} et~al.,}{{Martocchia}
  et~al.}{2018}]{martocchia18a}
{Martocchia} S.,  et~al., 2018, \mn@doi [\mnras] {10.1093/mnras/stx2556}, \href
  {http://adsabs.harvard.edu/abs/2018MNRAS.473.2688M} {473, 2688}

\bibitem[\protect\citeauthoryear{{McLaughlin} \& {van der Marel}}{{McLaughlin}
  \& {van der Marel}}{2005}]{mclaughlin05}
{McLaughlin} D.~E.,  {van der Marel} R.~P.,  2005, \mn@doi [\apjs]
  {10.1086/497429}, \href {http://adsabs.harvard.edu/abs/2005ApJS..161..304M}
  {161, 304}

\bibitem[\protect\citeauthoryear{{Mighell}, {Sarajedini}  \&
  {French}}{{Mighell} et~al.}{1998}]{mighell98}
{Mighell} K.~J.,  {Sarajedini} A.,   {French} R.~S.,  1998, \mn@doi [\aj]
  {10.1086/300591}, \href {http://adsabs.harvard.edu/abs/1998AJ....116.2395M}
  {116, 2395}

\bibitem[\protect\citeauthoryear{{Milone}, {Bedin}, {Piotto}  \&
  {Anderson}}{{Milone} et~al.}{2009}]{milone09}
{Milone} A.~P.,  {Bedin} L.~R.,  {Piotto} G.,   {Anderson} J.,  2009, \mn@doi
  [A\&A] {10.1051/0004-6361/200810870}, \href
  {http://adsabs.harvard.edu/abs/2009A%26A...497..755M} {497, 755}

\bibitem[\protect\citeauthoryear{{Milone} et~al.,}{{Milone}
  et~al.}{2012}]{milone12}
{Milone} A.~P.,  et~al., 2012, \mn@doi [\aap] {10.1051/0004-6361/201016384},
  \href {http://adsabs.harvard.edu/abs/2012A%26A...540A..16M} {540, A16}

\bibitem[\protect\citeauthoryear{{Milone} et~al.,}{{Milone}
  et~al.}{2017}]{milone17}
{Milone} A.~P.,  et~al., 2017, \mn@doi [\mnras] {10.1093/mnras/stw2531}, \href
  {http://adsabs.harvard.edu/abs/2017MNRAS.464.3636M} {464, 3636}

\bibitem[\protect\citeauthoryear{{Milone} et~al.,}{{Milone}
  et~al.}{2018}]{milone18}
{Milone} A.~P.,  et~al., 2018, preprint, \href
  {http://adsabs.harvard.edu/abs/2018arXiv180210538M} {} (\mn@eprint {arXiv}
  {1802.10538})

\bibitem[\protect\citeauthoryear{{Mucciarelli}, {Carretta}, {Origlia}  \&
  {Ferraro}}{{Mucciarelli} et~al.}{2008}]{mucciarelli08}
{Mucciarelli} A.,  {Carretta} E.,  {Origlia} L.,   {Ferraro} F.~R.,  2008,
  \mn@doi [AJ] {10.1088/0004-6256/136/1/375}, \href
  {http://adsabs.harvard.edu/abs/2008AJ....136..375M} {136, 375}

\bibitem[\protect\citeauthoryear{{Mucciarelli}, {Origlia}, {Ferraro}  \&
  {Pancino}}{{Mucciarelli} et~al.}{2009}]{mucciarelli09}
{Mucciarelli} A.,  {Origlia} L.,  {Ferraro} F.~R.,   {Pancino} E.,  2009,
  \mn@doi [ApJ] {10.1088/0004-637X/695/2/L134}, \href
  {http://adsabs.harvard.edu/abs/2009ApJ...695L.134M} {695, L134}

\bibitem[\protect\citeauthoryear{{Mucciarelli}, {Dalessandro}, {Ferraro},
  {Origlia}  \& {Lanzoni}}{{Mucciarelli} et~al.}{2014}]{mucciarelli14}
{Mucciarelli} A.,  {Dalessandro} E.,  {Ferraro} F.~R.,  {Origlia} L.,
  {Lanzoni} B.,  2014, \mn@doi [\apjl] {10.1088/2041-8205/793/1/L6}, \href
  {http://adsabs.harvard.edu/abs/2014ApJ...793L...6M} {793, L6}

\bibitem[\protect\citeauthoryear{{Muratov} \& {Gnedin}}{{Muratov} \&
  {Gnedin}}{2010}]{gnedin10}
{Muratov} A.~L.,  {Gnedin} O.~Y.,  2010, \mn@doi [ApJ]
  {10.1088/0004-637X/718/2/1266}, \href
  {http://adsabs.harvard.edu/abs/2010ApJ...718.1266M} {718, 1266}

\bibitem[\protect\citeauthoryear{{Nardiello} et~al.,}{{Nardiello}
  et~al.}{2018}]{nardiello18}
{Nardiello} D.,  et~al., 2018, \mn@doi [\mnras] {10.1093/mnras/sty2515}, \href
  {http://adsabs.harvard.edu/abs/2018MNRAS.481.3382N} {481, 3382}

\bibitem[\protect\citeauthoryear{{Niederhofer} et~al.,}{{Niederhofer}
  et~al.}{2017a}]{niederhofer17a}
{Niederhofer} F.,  et~al., 2017a, \mn@doi [MNRAS] {10.1093/mnras/stw2269},
  \href {http://adsabs.harvard.edu/abs/2017MNRAS.464...94N} {464, 94}

\bibitem[\protect\citeauthoryear{{Niederhofer} et~al.,}{{Niederhofer}
  et~al.}{2017b}]{niederhofer17b}
{Niederhofer} F.,  et~al., 2017b, \mn@doi [MNRAS] {10.1093/mnras/stw3084},
  \href {http://adsabs.harvard.edu/abs/2017MNRAS.465.4159N} {465, 4159}

\bibitem[\protect\citeauthoryear{{Parisi}, {Geisler}, {Clari{\'a}},
  {Villanova}, {Marcionni}, {Sarajedini}  \& {Grocholski}}{{Parisi}
  et~al.}{2015}]{parisi15}
{Parisi} M.~C.,  {Geisler} D.,  {Clari{\'a}} J.~J.,  {Villanova} S.,
  {Marcionni} N.,  {Sarajedini} A.,   {Grocholski} A.~J.,  2015, \mn@doi [\aj]
  {10.1088/0004-6256/149/5/154}, \href
  {http://adsabs.harvard.edu/abs/2015AJ....149..154P} {149, 154}

\bibitem[\protect\citeauthoryear{{Pietrinferni}, {Cassisi}, {Salaris}  \&
  {Castelli}}{{Pietrinferni} et~al.}{2004}]{pietrinferni04}
{Pietrinferni} A.,  {Cassisi} S.,  {Salaris} M.,   {Castelli} F.,  2004,
  \mn@doi [ApJ] {10.1086/422498}, \href
  {http://adsabs.harvard.edu/abs/2004ApJ...612..168P} {612, 168}

\bibitem[\protect\citeauthoryear{{Pietrzy{\'n}ski} et~al.,}{{Pietrzy{\'n}ski}
  et~al.}{2019}]{lmcdm}
{Pietrzy{\'n}ski} G.,  et~al., 2019, \mn@doi [\nat]
  {10.1038/s41586-019-0999-4}, \href
  {http://adsabs.harvard.edu/abs/2019Natur.567..200P} {567, 200}

\bibitem[\protect\citeauthoryear{{Piotto} et~al.,}{{Piotto}
  et~al.}{2015}]{piotto15}
{Piotto} G.,  et~al., 2015, \mn@doi [AJ] {10.1088/0004-6256/149/3/91}, \href
  {http://adsabs.harvard.edu/abs/2015AJ....149...91P} {149, 91}

\bibitem[\protect\citeauthoryear{{Renzini} et~al.,}{{Renzini}
  et~al.}{2015}]{renzini15}
{Renzini} A.,  et~al., 2015, \mn@doi [\mnras] {10.1093/mnras/stv2268}, \href
  {http://adsabs.harvard.edu/abs/2015MNRAS.454.4197R} {454, 4197}

\bibitem[\protect\citeauthoryear{{Rich}, {Shara}  \& {Zurek}}{{Rich}
  et~al.}{2001}]{rich01}
{Rich} R.~M.,  {Shara} M.~M.,   {Zurek} D.,  2001, \mn@doi [\aj]
  {10.1086/321164}, \href {http://adsabs.harvard.edu/abs/2001AJ....122..842R}
  {122, 842}

\bibitem[\protect\citeauthoryear{{Schiavon}, {Caldwell}, {Conroy}, {Graves},
  {Strader}, {MacArthur}, {Courteau}  \& {Harding}}{{Schiavon}
  et~al.}{2013}]{schiavon13}
{Schiavon} R.~P.,  {Caldwell} N.,  {Conroy} C.,  {Graves} G.~J.,  {Strader} J.,
   {MacArthur} L.~A.,  {Courteau} S.,   {Harding} P.,  2013, \mn@doi [\apjl]
  {10.1088/2041-8205/776/1/L7}, \href
  {http://adsabs.harvard.edu/abs/2013ApJ...776L...7S} {776, L7}

\bibitem[\protect\citeauthoryear{{Stetson}}{{Stetson}}{1987}]{stetson87}
{Stetson} P.~B.,  1987, \mn@doi [PASP] {10.1086/131977}, \href
  {http://adsabs.harvard.edu/abs/1987PASP...99..191S} {99, 191}

\bibitem[\protect\citeauthoryear{{Stetson}}{{Stetson}}{1994}]{stetson94}
{Stetson} P.~B.,  1994, \mn@doi [PASP] {10.1086/133378}, \href
  {http://adsabs.harvard.edu/abs/1994PASP..106..250S} {106, 250}

\bibitem[\protect\citeauthoryear{{Villanova}, {Geisler}, {Carraro}, {Moni
  Bidin}  \& {Mu{\~n}oz}}{{Villanova} et~al.}{2013}]{villanova13}
{Villanova} S.,  {Geisler} D.,  {Carraro} G.,  {Moni Bidin} C.,   {Mu{\~n}oz}
  C.,  2013, \mn@doi [ApJ] {10.1088/0004-637X/778/2/186}, \href
  {http://adsabs.harvard.edu/abs/2013ApJ...778..186V} {778, 186}

\bibitem[\protect\citeauthoryear{{Walker} et~al.,}{{Walker}
  et~al.}{2011}]{walker11}
{Walker} A.~R.,  et~al., 2011, \mn@doi [MNRAS]
  {10.1111/j.1365-2966.2011.18736.x}, \href
  {http://adsabs.harvard.edu/abs/2011MNRAS.415..643W} {415, 643}

\bibitem[\protect\citeauthoryear{{van Dokkum}}{{van
  Dokkum}}{2001}]{vandokkum01}
{van Dokkum} P.~G.,  2001, \mn@doi [PASP] {10.1086/323894}, \href
  {http://adsabs.harvard.edu/abs/2001PASP..113.1420V} {113, 1420}

\makeatother
\end{thebibliography}

%%%%%%%%%%%%%%%%%%%%%%%%%%%%%%%%%%%%%%%%%%%%%%%%%%

%%%%%%%%%%%%%%%%% APPENDICES %%%%%%%%%%%%%%%%%%%%%

%\appendix

%\section{Some extra material}

%%%%%%%%%%%%%%%%%%%%%%%%%%%%%%%%%%%%%%%%%%%%%%%%%%

% Don't change these lines
\bsp	% typesetting comment
\label{lastpage}

%\clearpage\pagestyle{empty}
%
%\addtolength{\textheight}{0.65in}

\end{document}